\documentclass[preprint2]{aastex}

\usepackage{natbib}
\usepackage{rotating}
\textheight=\baselinestretch\textheight
\ifdim\textheight>25.2cm\textheight=25.0cm\fi

\topskip\baselineskip
\maxdepth\baselineskip
\let\tighten=\relax
\let\tightenlines=\tighten

\newcommand{\kms}{\mbox{km\,s$^{-1}$}}

\slugcomment{To appear in ApJ Supplement.}

\shortauthors{[V.A. Moss et al.}

\begin{document}

\title{High velocity clouds in the Galactic All Sky Survey I. Catalogue} 

\author{
V.~A.\,Moss\altaffilmark{1,2,}$^{\dagger}$,
N.~M.\,McClure-Griffiths\altaffilmark{2},
T.\,Murphy\altaffilmark{2,3},
D.~J.\,Pisano\altaffilmark{4,5},
J.~K.\,Kummerfeld\altaffilmark{1},
and J.~R.\,Curran\altaffilmark{3}
}

\altaffiltext{$\dagger$}{\bf vmoss@physics.usyd.edu.au}
\altaffiltext{1}{Sydney Institute for Astronomy, School of Physics A29, University of Sydney, NSW 2006, Australia}
\altaffiltext{2}{CSIRO Astronomy and Space Science, ATNF, PO Box 76, Epping, NSW 1710, Australia}
\altaffiltext{3}{School of Information Technologies, University of Sydney, NSW 2006, Australia}
\altaffiltext{4}{Department of Physics, West Virginia University, PO Box 6315, Morgantown, WV 26506}
\altaffiltext{5}{Adjunct Assistant Astronomer at National Radio Astronomy Observatory, PO Box 2, Green Bank, WV 24944}

\begin{abstract}
We present a catalogue of high-velocity clouds (HVCs) from the Galactic All Sky Survey (GASS) of southern-sky neutral hydrogen, which has 57 mK sensitivity and 1 km s$^{-1}$ velocity resolution and was obtained with the Parkes Telescope. Our catalogue has been derived from the stray-radiation corrected second release of GASS. We describe the data and our method of identifying HVCs and analyse the overall properties of the GASS population. We catalogue a total of 1693 HVCs at declinations $< 0^\circ$, including 1111 positive velocity HVCs and 582 negative velocity HVCs. Our catalogue also includes 295 anomalous velocity clouds (AVCs). The cloud line-widths of our HVC population have a median FWHM of $\sim$19~\kms, which is lower than found in previous surveys. The completeness of our catalogue is above 95\% based on comparison with the HIPASS catalogue of HVCs, upon which we improve with an order of magnitude in spectral resolution. We find 758 new HVCs and AVCs with no HIPASS counterpart. The GASS catalogue will shed an unprecedented light on the distribution and kinematic structure of southern-sky HVCs, as well as delve further into the cloud populations that make up the anomalous velocity gas of the Milky Way.
\end{abstract}



\section{Introduction}
Fifty years after the discovery of high-velocity neutral hydrogen (H{\sc i}) in the halo surrounding the Milky Way \citep{Muller:1963p22387}, current research still seeks to understand the nature, origins and interactions of this anomalous gas which does not follow normal Galactic rotation. It is well-known that the most dense forms of high-velocity H{\sc i} occur as clouds organised into associations and complexes \citep{Wakker:1991p22252}. It is clear that their physical origins can be diverse, with evidence for complexes both extragalactic \citep{Wakker:2002p22578,Sembach:2002p22598,Lehner:2009p22600,Tripp:2012p22503} and Galactic \citep{Lehner:2010p22602,Smoker:2011p22447} in origin. These studies trace origins for extragalactic clouds based on their low metallicities, lack of molecular gas or association with the Magellanic system, and origins for Galactic clouds based on their ionisation content and low upper distance limits from stellar absorption lines. In northern and southern sky surveys of high-velocity clouds (HVCs) the Magellanic system (consisting of the Magellanic Clouds, the Magellanic Stream and the Leading Arm) contributes a large fraction of clouds, forming a prime example of the interaction between the Milky Way and extragalactic gas. But it is also clear that a considerable amount of anomalous velocity gas appears to be associated with interactions between the disk and the halo of the Milky Way and may in fact be of Galactic origin. High velocity gas has been discovered in other nearby galaxies such as M101, M31, M33 and NGC 205 \citep{TenorioTagle:1988p13608,Westmeier:2005p23593,Westmeier:2007p23592} and may be related to extraplanar gas observed in more distant galaxies \citep{Sancisi:2008p23612,Kamphuis:2011p23605}, with the origins of the gas indeterminate in many cases but most likely a combination of galactic activity and accretion of infalling gas to varying proportions depending on the galaxy \citep{Boomsma:2008p23611,Heald:2011p23622}.

The first surveys of Galactic HVCs focused on the northern hemisphere, with clouds identified to be inhomogeneous and hence grouped in complexes on the basis of their spatial and spectral proximity to other clouds \citep{Wannier:1972p23447,Giovanelli:1973p23469,Hulsbosch:1978p23464,Wakker:1991p22252}. Since then, many new surveys of H{\sc i} with increasing sensitivity and resolution have been released in both northern and southern hemispheres, including the Leiden-Dwingeloo survey \citep{Hartmann:1997p10021},  the Leiden-Argentina-Bonn (LAB) survey \citep{Kalberla:2005p22278} and the GALFA-HI survey \citep{Peek:2012p22763}. Each new study of anomalous velocity gas at higher angular resolution or higher spectral resolution reveals different aspects of the nature of this gas, resulting in new classes of anomalous velocity clouds (AVCs) such as compact HVCs \citep{Braun:1999p22713}, ultra-compact HVCs \citep{Bruns:2004p20402,Giovanelli:2010p23648}, projection-affected low velocity halo clouds \citep{Peek:2009p16944} and cold/warm low-velocity clouds \citep{Saul:2012p22666}.

The high spectral resolution of new H{\sc i} surveys allows us to examine the velocity structure of HVCs, which probes their physical conditions and interactions with their surroundings. Based on northern hemisphere surveys it has been found that the cloud line-widths of HVCs have a median full-width at half-maximum (FWHM) of $\sim$20-30~km~s$^{-1}$ \citep{Cram:1976p22750,deHeij:2002p12239,Kalberla:2006p22170}, corresponding to a kinetic temperature of $\sim$10$^4$~K. The nature of these clouds of H{\sc i} as relatively cold concentrations of gas means that their existence in a hot $\sim$10$^{6}$~K halo results in clear evidence of interaction. The form of the pressure and density gradients as a function of scale height position in the Milky Way is also thought to play a key role in the confinement and lifetime of these clouds. Based on the influences of temperature and pressure, the complex spatial and spectral structure of HVCs can provide clues about the physical properties and role of their surrounding environment in their evolution and dynamics.

We use data from the southern Galactic All Sky Survey (GASS) in H{\sc i}, obtained using the Parkes 64-m radio telescope \citep{McClureGriffiths:2009p3462}. GASS has excellent image fidelity due to its frequency-switching observing mode, and as such recovers emission on all angular scales. The second release of GASS (GASS~II) features improved sensitivity and stray-radiation correction \citep{Kalberla:2010p19880}, on which we performed automated source finding to identify HVCs. The catalogue most closely related to ours is the HIPASS HVC catalogue \citep{Putman:2002p12244}, which shares similar angular resolution and sky coverage but differs in spectral resolution and sensitivity. HIPASS was originally intended as an extragalactic survey, and as such the data used for constructing the HIPASS HVC catalogue had a low velocity resolution of 26.4~km~s$^{-1}$ (after initial smoothing) which provides a useful comparison to the GASS data with its spectral resolution of $\sim$1~km~s$^{-1}$. HIPASS also differs from GASS in its better sensitivity of 8~mK compared with the $\sim$40~mK RMS sensitivity of GASS~II. Taking into account both the effects of spectral resolution and sensitivity means that the two surveys are comparable to each other, though each is sensitive to different  characteristics of the HVC population. HIPASS was also taken in a different observing mode, and as such even with the MINMED5 data reduction technique had limited sensitivity to emission on scales $>$ 7$^\circ$.

This paper presents a catalogue of HVCs and AVCs (anomalous velocity clouds which do not meet the traditional velocity criteria of high velocity gas but deviate significantly from Galactic rotation) compiled from the GASS~II data. In Section \ref{methodology}, we give details of the data and source finding procedure. Section \ref{results} presents our catalogue of GASS HVCs as well as the overall population characteristics. We consider the completeness and reliability of our catalogue in Section \ref{completeness}, in the context of comparison with the nearest similar catalogue of HIPASS HVCs. We conclude by summarising our catalogue and giving an indication of our directions for future work using the GASS HVC catalogue.


\section{The Galactic All-Sky Survey II: data and methodology}\label{methodology}
Our catalogue is derived from the second release of GASS, which surveyed southern-sky H{\sc i} \citep{McClureGriffiths:2009p3462,Kalberla:2010p19880}, covering declinations south of $\delta$ = 1$^\circ$. We have used all GASS data south of 0$^\circ$ in the construction of our survey in order to avoid regions of increased noise near the limits of the survey and to limit our catalogue to a southern-sky survey. GASS was undertaken with the Parkes 64-m radio telescope using the 21~cm multibeam and consisted of two sets of scans, one in right ascension and one in declination. The effective angular resolution of GASS is $\sim$16$'$ with a spectral resolution of 1~km~s$^{-1}$ and $\sim$57 mK RMS noise. As GASS focuses specifically on the H{\sc i} content of the Milky Way, the survey velocity range is complete from approximately $-468$~km~s$^{-1}$ to $+468$~km~s$^{-1}$. GASS used in-band frequency-switching to assist bandpass correction while maximising the observing efficiency and maintaining sensitivity to large-scale structure. This involved recording spectra at two spectrally-close frequencies, which allows removal of continuum and time-varying system gains, as well as maximising on-source time by combining the frequency-switched pairs to create continuous bandwidth. Frequency switching in this way means that a negative copy of a spectral line will be reproduced at $\pm660$~km~s$^{-1}$, which is important for consideration of artifacts. The second release differs from the first with the further reduction of artifacts and RFI, as well as correction of stray radiation artifacts (arising from leakage through the sidelobes of a radio telescope due to an imperfect beam) using the Leiden/Argentina/Bonn (LAB) survey \citep{Kalberla:2005p22278}. The RMS noise of GASS~II exhibits dependence on Galactic latitude, ranging from 20 mK to 60 mK depending on proximity to bright emission \citep{Kalberla:2010p19880}. In creating the GASS catalogue of HVCs, it was important to have a global sensitivity across the southern sky (particularly to be consistent with the RMS noise near the Galactic Plane for source finding), and so we adopted the originally published noise of 57~mK to guarantee reliable detections throughout the data. 

\begin{figure*}
  \centering
  \includegraphics[width=0.35\textwidth, angle=-90, trim=0 0 0 0]{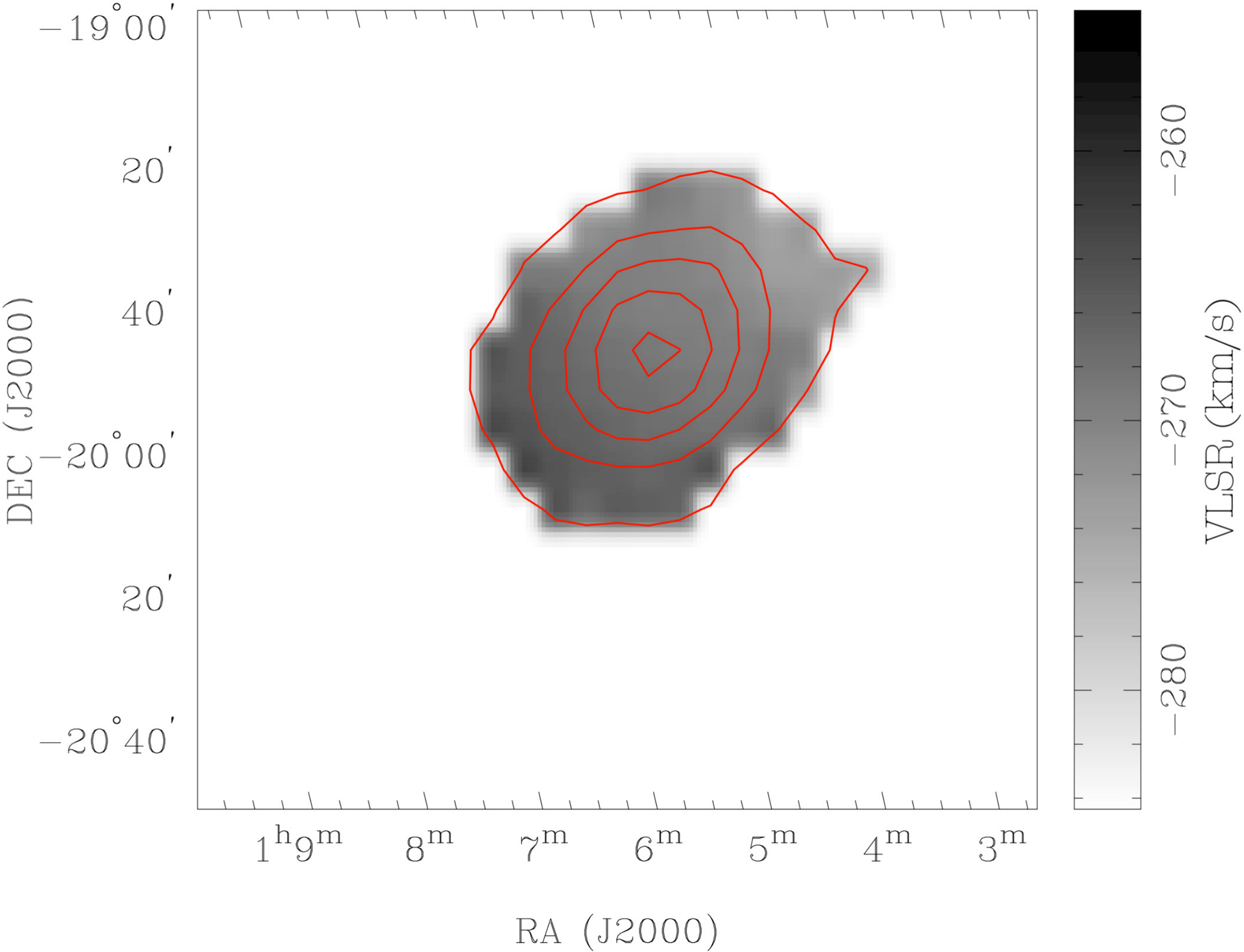}
  \includegraphics[width=0.35\textwidth, angle=-90, trim=0 0 0 0]{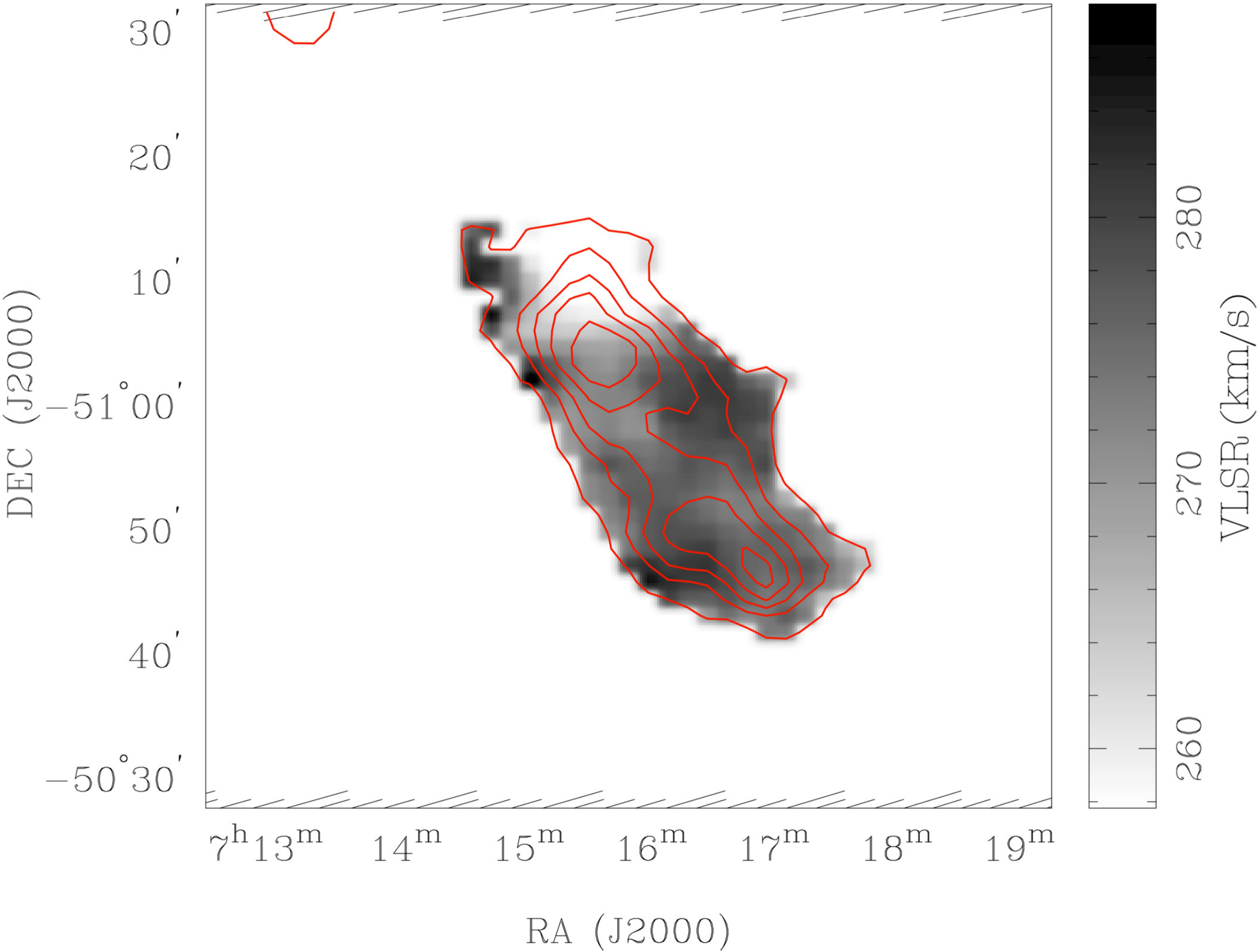}
   \includegraphics[width=0.35\textwidth, angle=-90, trim=0 0 0 0]{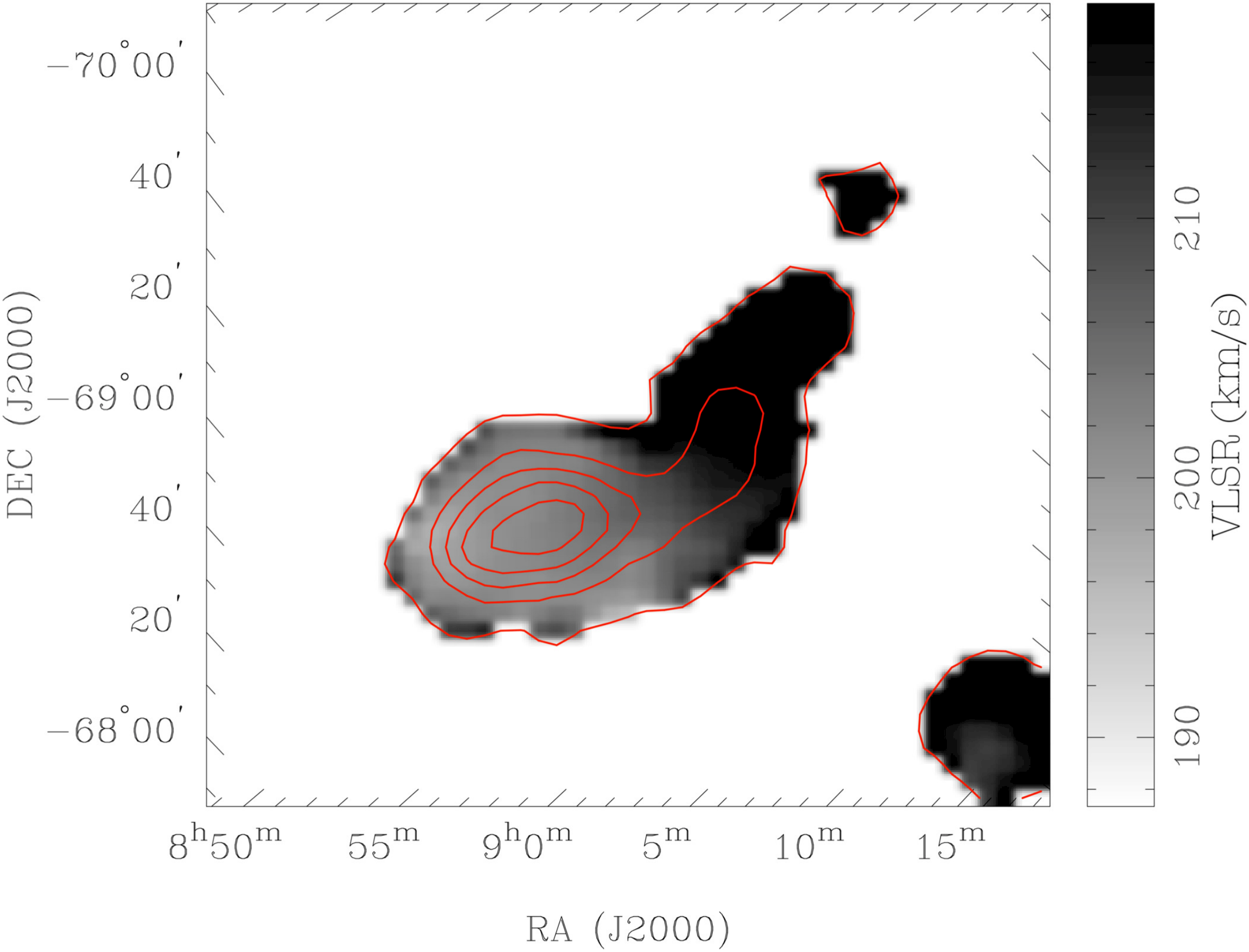}
  \caption{\small Examples of the different types of clouds evident in the GASS data. Each cloud is an example of an isolated (top left), spatial multiple-component (top right) or head-tail (bottom) cloud. The background image is the 1st-moment map showing the velocity structure of the source, with 0th-moment contours showing intensity structure in red ranging from $\sigma$ to 90\% of the integrated intensity peak (corresponding to 28.2, 10.7 and 58.2~K~km~s$^{-1}$) in 5 equal steps, where $\sigma$ is the GASS noise of 57~mK multiplied by 2 $\times$ the spectral FWHM of each source.}
  \label{fig:class}
\end{figure*}

\subsection{Source finding methodology}\label{method}
To identify HVCs, we use an automated source finding routine which adopts the flood fill algorithm. The flood fill algorithm searches for emission connected from a peak down to a user-specified threshold, and identifies this as an object or a source. We specifically adopt the algorithm developed and used by \citet{Murphy2007a}, but extended to a three-dimensional cube. The flood fill algorithm works in the following way: 
\begin{enumerate}
\item Identify all peaks $>$ the threshold peak level ($T_{pk}$).
\item Extend from the peak in each dimension (in our case: $\alpha$, $\delta$, $v$), amalgamating pixels into one source until the cutoff level ($T_{min}$) is reached.
\item Record this as a candidate source. 
\end{enumerate}
We give examples of the kinds of clouds identified by our source finder in Figure \ref{fig:class}. In order to reduce the impact of noise and artifacts, we searched on cubes binned by averaging 5 channels in the spectral axis (see Section \ref{binning} for more details about binning). We set a minimum source width of 3 pixels in both spatial directions, which is close to the size of the GASS beam in the spatial axes, and to 3 binned spectral channels (minimum total width of $\sim$12 km s$^{-1}$). For our flood fill parameters, we choose a peak of $T_{pk} = 4\sigma$ and $T_{min} = 2\sigma$, extrapolating from the GASS noise of 57~mK to the binned cubes. As stated above, we adopted the RMS noise of the first release of GASS to guarantee a consistent catalogue sensitivity across the sky and near the Galactic plane, although the noise of GASS~II is lower at higher Galactic latitudes. The nominal column density limit for a single channel at this noise of 57~mK is $\sim$8.5~$\times$~10$^{16}$~cm$^{-2}$. 

\begin{figure*}
  \centering
  \includegraphics[width=0.8\textwidth, angle=0, trim=0 0 0 0]{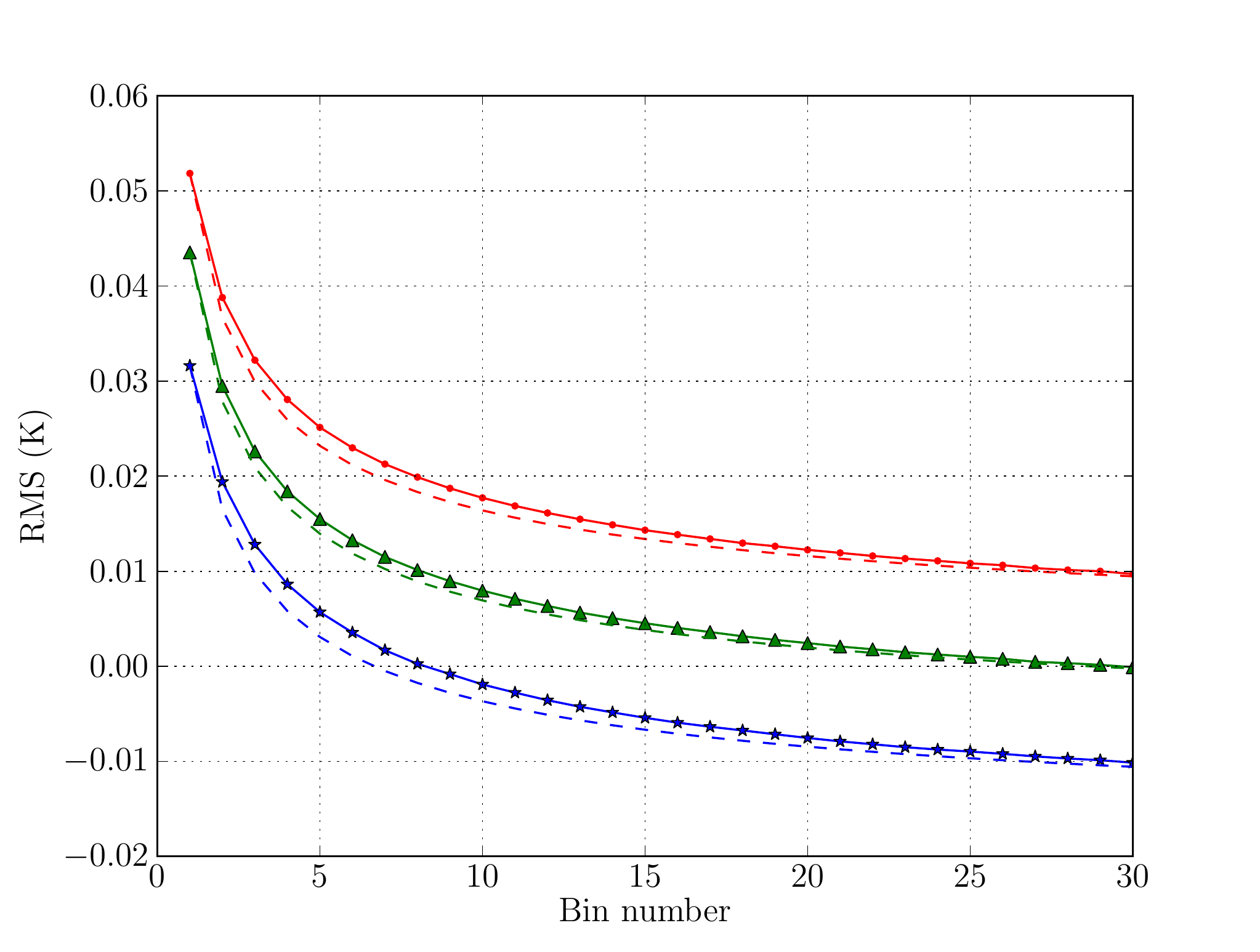}
  \caption{\small The results of spectrally binning three independent ``noise'' cubes up to 30 channels are shown. Each colour represents the outcome from a different cube, offset in the top figure by 0.00~K (red, circles), -0.01~K (green, triangles) and -0.02~K (blue, stars) in order to make the behaviour of the individual lines clear. The dashed line in each case represents the expected theoretical trend following Poisson statistics. The data closely follow the theoretical noise. The difference between the RMS in the data and the theoretical RMS peaks at 3 bins for all three noise cubes, with a maximum difference of 3~mK.}
  \label{fig:noise}
\end{figure*}

We examined the effect of binning using three cubes that contain predominantly noise, calculating the change in RMS with increasing number of bins. The results are shown in Figure \ref{fig:noise}. We find that the GASS noise closely follows the theoretically expected trend with increasing bin number (based on Poisson statistics), although it is slightly higher than the theoretical noise. The difference across these three samples of noise peaks at a value of at most 3~mK in the case of binning by 3 bins and occurs at $\sim$2.6~mK for the case of 5 bins. This is a very small difference overall and simply results in probing further into the noise by assuming the theoretical limit. We thus assume the N-binned RMS, $\sigma_{N}$, to be $\sim$$\sigma N^{-1/2}$, and calculate $\sigma_5 \sim 25$ mK, which gives $2\sigma_5 \sim 50$ mK and $4\sigma_5 \sim 100$ mK. We use these values as our $T_{min}$ and $T_{pk}$ respectively. For comparison to the data in our noise cubes, we found a 4$\sigma$ 5-binned RMS of 101, 102 and 106~mK in each which is very close to the theoretical 100~mK value. Although a $T_{pk}$ of $4\sigma$ is lower than most automated source finding, the combination of this with the 3 channel minimum in each axis increases the significance of the detections.

\begin{figure*}
  \centering
  \includegraphics[width=0.49\textwidth, angle=0, trim=0 0 0 0]{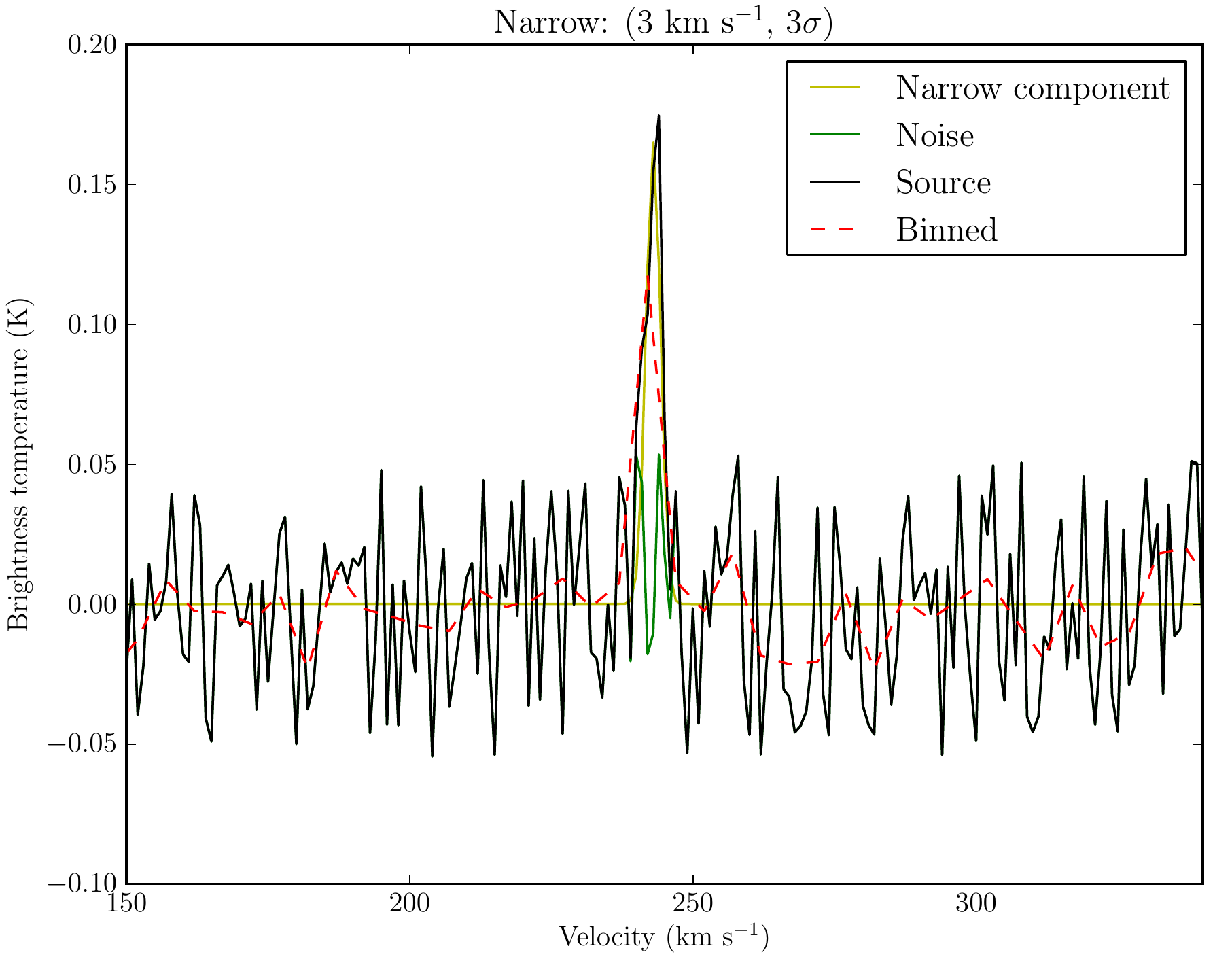}
    \includegraphics[width=0.49\textwidth, angle=0, trim=0 0 0 0]{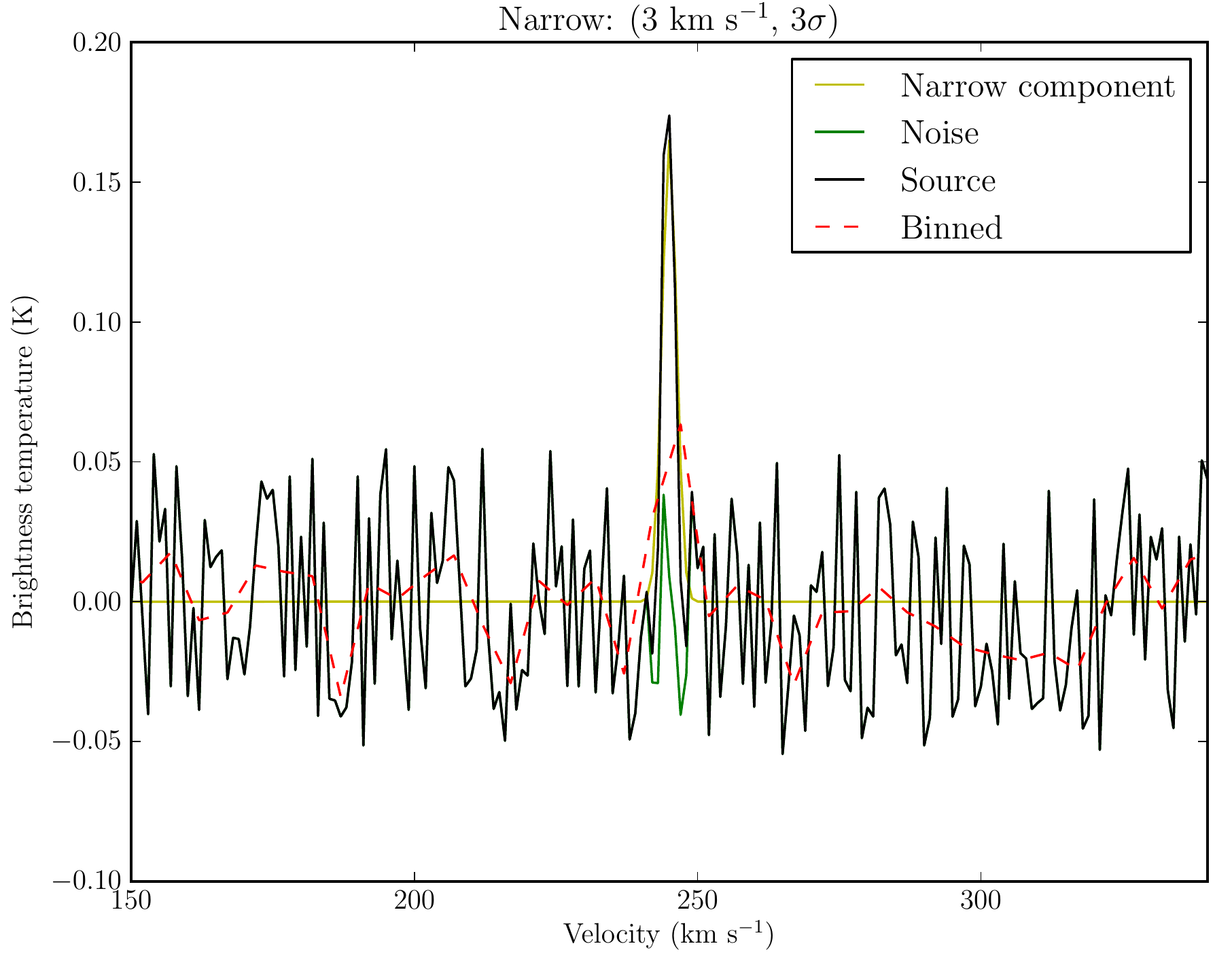}
  \caption{\small Model narrow-line clouds of peak brightness 3$\sigma$ and FWHM of 3~km~s$^{-1}$, falling within a spectral bin (left) or on a spectral bin edge (right) during the binning process. The red dashed line shows the resultant spectrum after binning each cloud by 5 channels. These plots show the potentially detrimental effects of binning on low-brightness, low-width clouds. For such a narrow and low signal-to-noise cloud, the effect of its position in the spectrum during binning is notable and significantly reduces its detectability if the cloud falls on the edge of a bin. The numbers in brackets show the FWHM and brightness of the simulated cloud.}
  \label{fig:nodeantinode}
\end{figure*}

\subsection{Data processing}\label{dataproc}
One of the challenges of identifying HVCs close to Galactic velocities is distinguishing them from emission that is due solely to Galactic rotation near the centre of the Milky Way. We use the parameter of deviation velocity, which indicates how much the object's velocity deviates from a simple model of Galactic rotation where $v_{dev}$ = $v_{LSR} - v_{model}$. As HVCs can be considered to require a deviation velocity $v_{dev}$ of at least 50~km~s$^{-1}$ \citep{Wakker:1991p22250,Wakker:1997p9859}, we use this definition to help separate HVCs from high-velocity Milky Way gas. To determine $v_{dev}$, we construct a model of the Galactic H{\sc i} distribution.  We have chosen to use the model of de Heij et al. (2002) to facilitate comparison with past studies of HVCs and intermediate velocity clouds (IVCs).  This model is composed of a thin disk with a constant central density in H{\sc i}, $n_{0}$~=~0.35 cm$^{-3}$, and kinetic  temperature, $T_K$~=~100~K, within Galactocentric radius $R$~=~11.5~kpc.  Beyond 11.5~kpc, the midplane density in H{\sc i} is
\begin{equation}
n =n_{0} \exp({-\frac{R-11.5~{\rm kpc}}{3~{\rm kpc}}}).
\end{equation}
In the vertical direction, the H{\sc i} density follows a Gaussian distribution with
\begin{equation}
n =n_0 \exp ({-\frac{(z-z_0)^2}{2\sigma_z^2}}),
\end{equation}
where $z$ is the scale height of the Milky Way in pc. For R $<$ 11.5~kpc, $\sigma_z$~=~180~pc, and $z_0$~=~0~pc. For $R$ $>$ 11.5~kpc, $\sigma_z  = 180 + 80 (R - 11.5$ kpc) pc, and $z_0$~=~($R$~$-$~11.5)/6~sin$\phi$ + ($R$~$-$~11.5/6)$^2$(1~$-$~2~cos$\phi$)~pc.  $\phi$ is the cylindrical Galactic azimuth (increasing in the direction of Galactic rotation) centred on the Galactic centre where $\phi$~=~180$\arcdeg$ is the direction toward the Sun.  The variation of $z_0$ and $\sigma_z$ beyond $R$~=~11.5~kpc provides for the
warp and flare of the Galactic disk.  The Milky Way is assumed to have a flat rotation curve with $R_0$~=~8.5~kpc, $v_{rot}$~=~220~\kms, and
$\sigma_v$~=~20~\kms.  

We constructed a model using a 50~pc grid out to a Galactocentric radius of 50~kpc where

\begin{equation}
T_B(v)=T_K (1-e^{-\tau_v}),
\end{equation}
and where
\begin{equation}
\tau_v = \frac{33.52 n_0}{T_K \sigma_v} \exp ({-\frac{v-v_r}{2 \sigma_v^2}}),
\end{equation}
with all temperatures in K, velocities in \kms\ and densities in cm$^{-3}$.  

Using this model, we identify the most positive and negative $v_{LSR}$ for H{\sc i} in the model with $T_B \ge$ 0.5 K at every position in the sky; 
these velocities are the basis for calculating $v_{dev}$. The collective anomalous velocity gas deviating from Galactic rotation in the Milky Way is typically categorised as low velocity gas, intermediate velocity gas or high velocity gas depending on $v_{LSR}$, $v_{dev}$ or a combination of the two. This had led to the existence and studies of such objects as low velocity clouds (LVCs) and intermediate velocity clouds (IVCs) as well as HVCs, which makes the physical distinction between each class of anomalous velocity cloud more important as there is a significant degree of overlap between the three groups of anomalous velocity gas. In order to guarantee detection of cloud emission peaking at deviation velocities of 50~km~s$^{-1}$ in accordance with the deviation velocity cutoff proposed by \citet{Wakker:1991p22250} and in this context of anomalous velocity clouds, we mask out all emission in our data that has a deviation velocity of $< 30$ km s$^{-1}$. We refer to any GASS clouds which do not meet the HVC velocity criteria (see Section \ref{results}) as anomalous velocity clouds (AVCs) because they exist on the blurry boundary between IVC and HVC, and as such we refrain from classifying them in absence of further investigation into their physical properties.

Due to the masking process, as well as the effects of frequency-switching, we found it necessary to manually remove any sources which appeared to either be poorly masked Galactic emission, artifacts near regions of `ghosting' where some of the frequency-shifted emission is incompletely subtracted or regions of generally higher noise. The initial source finding routine resulted in $\sim$6000 candidate sources, which were then inspected to ensure the reliability of the final catalogue. This inspection reduced the total number of clouds to $\sim$2000. The decision on each source was made in the context of all information available, including measured properties, spectra, moment maps and proximity to artifact regions.

\begin{figure*}
  \centering
  \includegraphics[width=0.4\textwidth, angle=-90, trim=0 0 0 0]{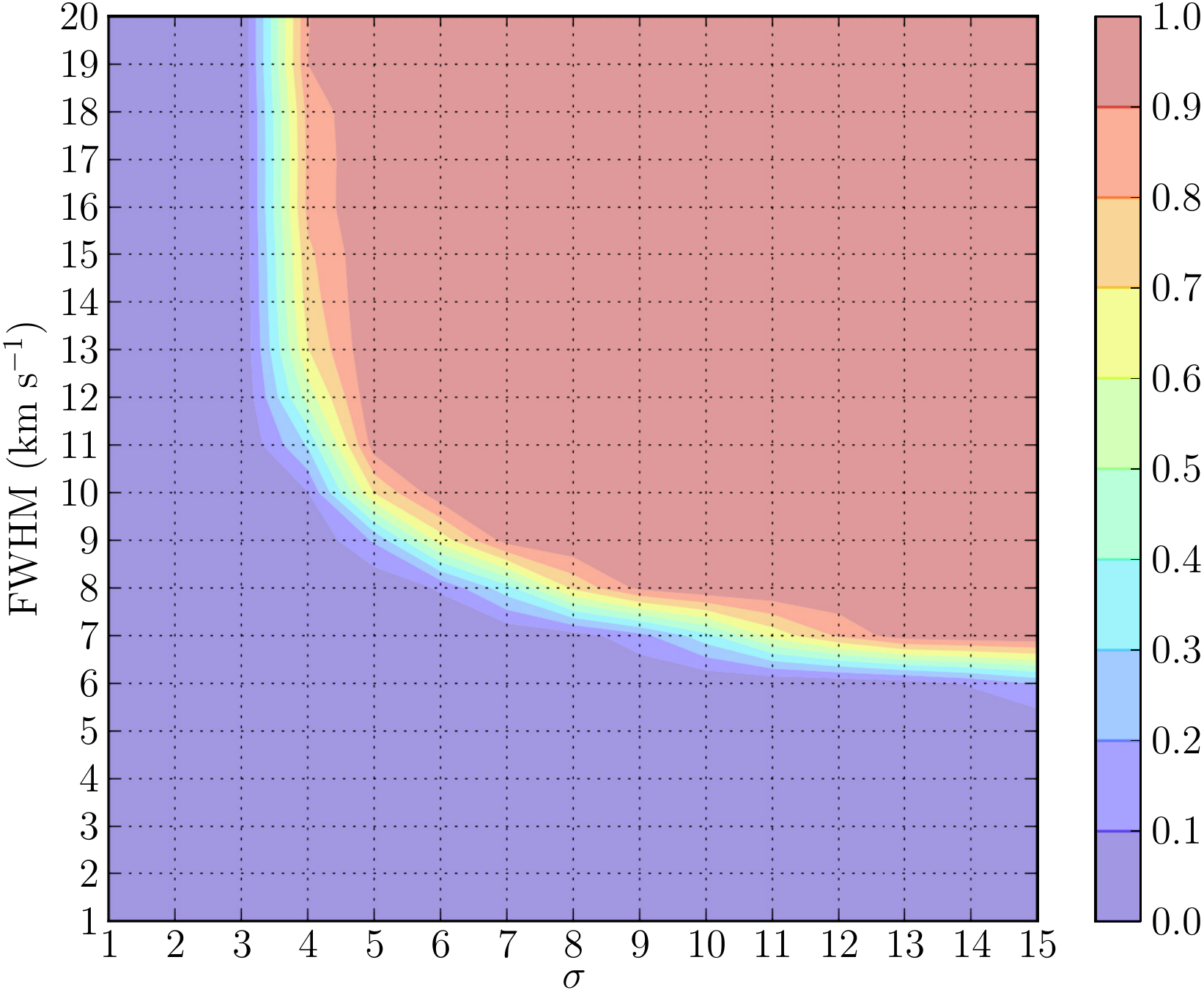}
    \includegraphics[width=0.4\textwidth, angle=-90, trim=0 0 0 0]{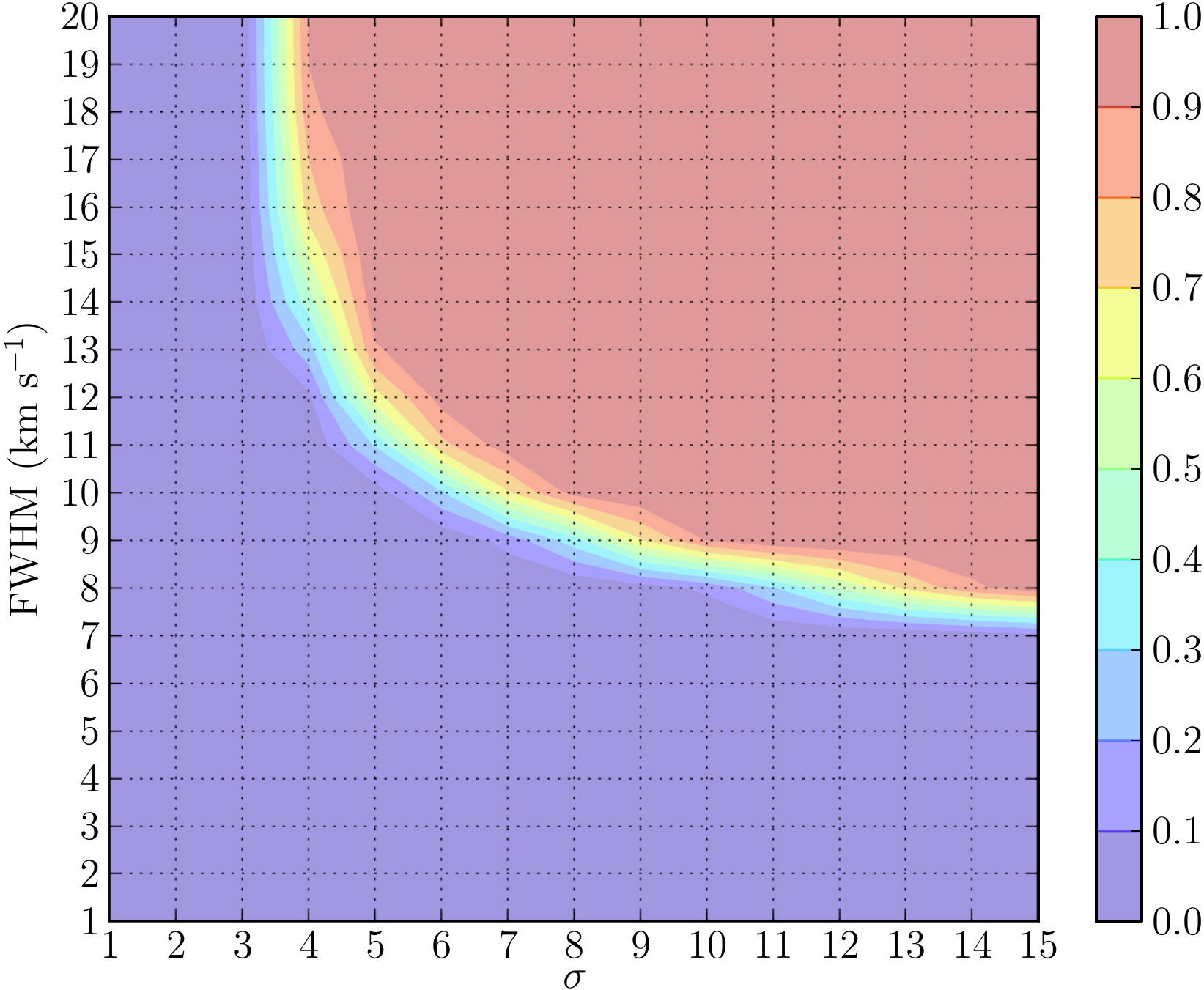}
  \caption{\small Detectability plots for model narrow-line clouds binned by 5 channels falling bin within a spectral  (left) and on the edge of a spectral bin (right). The red regions represent a detectability of 90\% or greater and can be used as a completeness threshold for these clouds. We see that even for narrow-line clouds of 15$\sigma$ brightness, we expect to be limited to detecting clouds of greater than 7 or 8~km~s$^{-1}$ FWHM (depending on whether they fall within a spectral bin or on the edge of a spectral bin). Conversely, at the largest FWHM of 20~km~s$^{-1}$, we are likely to be complete down to $\sim$4$\sigma$ regardless of where they peak.}
  \label{fig:detectability}
\end{figure*}

\begin{figure*}
  \centering
  \includegraphics[width=0.4\textwidth, angle=-90, trim=0 0 0 0]{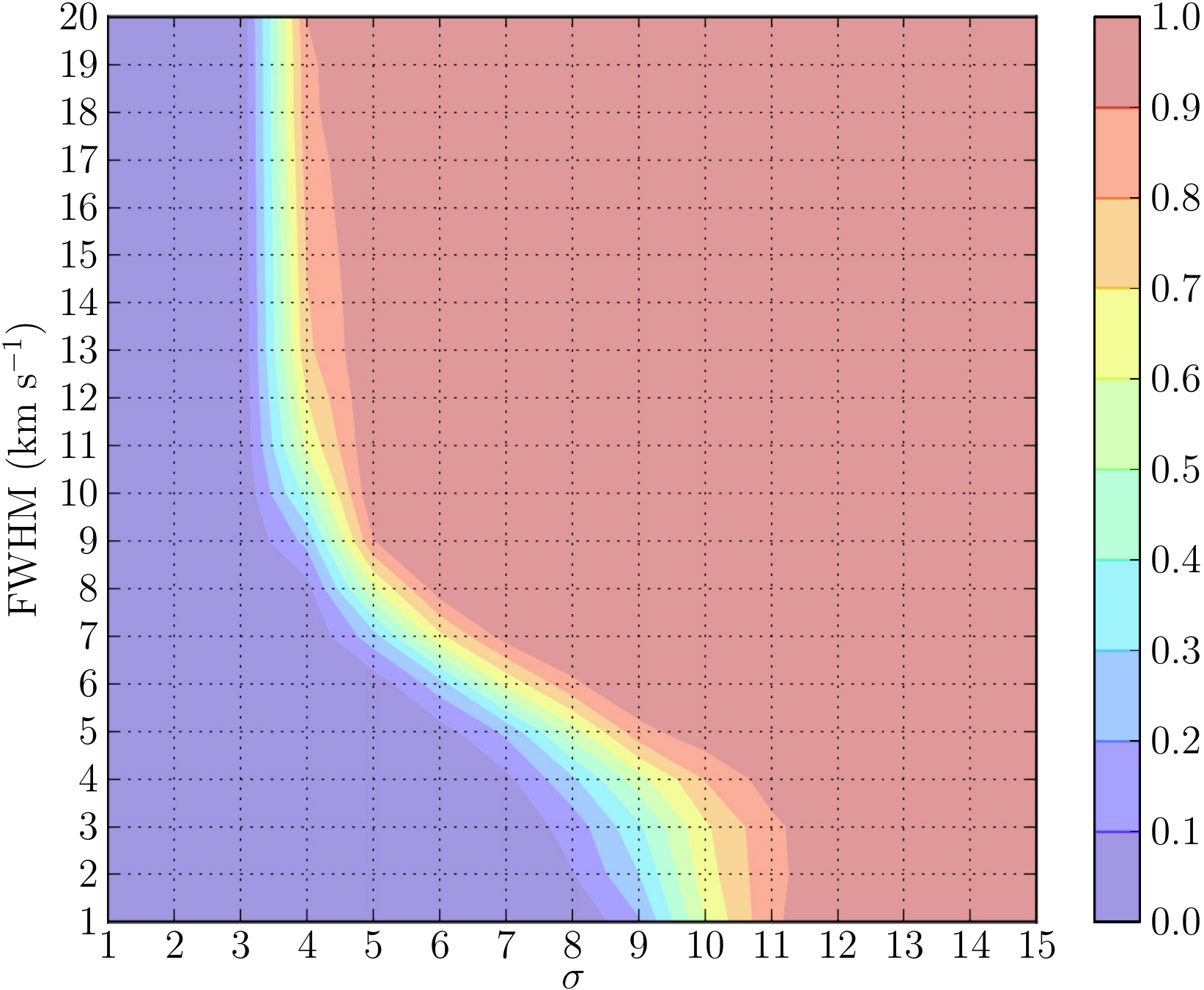}
    \includegraphics[width=0.4\textwidth, angle=-90, trim=0 0 0 0]{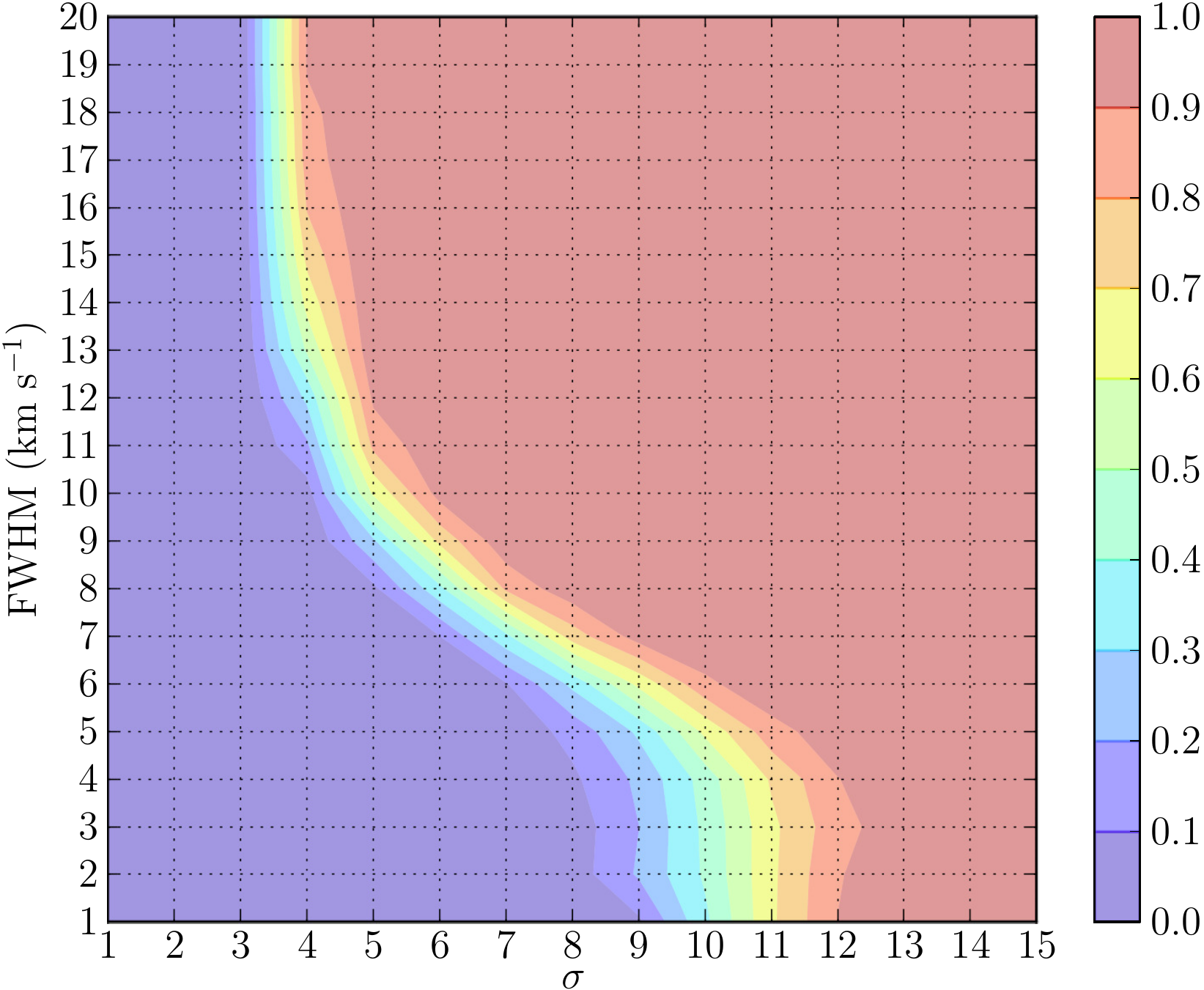}
  \caption{\small Detectability plots for model narrow-line clouds embedded within a broad component of FWHM 20~km~s$^{-1}$ binned by 5 channels falling within a spectral bin (left) and on the edge of a spectral bin (right). The red regions represent a detectability of 90\% or greater and can be used as a completeness threshold for these clouds. In this case, we see that we expect to be complete even down to a FWHM of 1~km~s$^{-1}$ for clouds brighter than $\sim$12$\sigma$. The brightness temperature $\sigma$ here refers to the combined brightness of both the narrow and the broad component.}
  \label{fig:detectability2}
\end{figure*}

\subsection{Effect of binning}\label{binning}
To investigate the effect of the spectral binning on our ability to find narrow spectral line (narrow-line) clouds, we simulated the binning process on narrow-line clouds of varying signal-to-noise. We did this by generating model clouds of varying FWHM and varying peak brightness temperature embedded in Gaussian noise of 57~mK. We then performed binning on the spectrum of each cloud and measured the resulting properties of the binned spectrum. We considered both the best case scenario (a narrow cloud peaks within the edges of a spectral bin) and the worst case scenario (a narrow cloud peaks on the edge of a spectral bin) in order to understand at what level we are sensitive to narrow-line clouds when binning the data by 5 channels. The effect of a cloud's position within a spectral bin or on a bin edge is only really apparent for low brightness, very narrow clouds. This is shown in Figure \ref{fig:nodeantinode} for a 3$\sigma$ cloud with a FWHM of 3~km~s$^{-1}$, where the red dashed line represents the binned spectrum. The detrimental effect on peak flux of falling on the edge of a bin is very clear for this type of cloud.

We generated model clouds over the peak brightness temperature range of [1,~15]~$\sigma$ in steps of 1$\sigma$ and over the FWHM range of [1,~20]~km~s$^{-1}$ in steps of 1~km~s$^{-1}$ both within a spectral bin and on a bin edge. We define a cloud as detectable if, in the binned spectrum, the peak is greater than our required source finding peak (100~mK) and if the positions on either side of the peak are greater than our required source finding threshold (50~mK). We simulated each cloud at each FWHM and $\sigma$ 1000 times to average out the effect of statistical noise, and kept track of how many clouds in each case are `detected'. From this we generated a detectability map for clouds. This is shown in Figure \ref{fig:detectability}. We see that we are likely to miss clouds of width less than $\sim$8~km~s$^{-1}$ even at the maximum simulated brightness temperature of 15$\sigma$ in our binned search.

We compared this result to two-component clouds, where narrow-line components are embedded within a broad component of FWHM comparable to that of the median HVC population ($\sim$20~km~s$^{-1}$) \citep{BenBekhti:2006p22800,Kalberla:2006p22170}. We performed the same simulations as before, except this time we added a broad component at the same position as the narrow-line component with one-third of its intensity such that the total brightness is at the simulated $\sigma$ level. The detectability plots are shown in Figure \ref{fig:detectability2}. We see that the embedding of a narrow-line component within a broad component, even if it is faint, makes a considerable difference to the detectability of the source in total. We are likely to be complete down to the narrowest FWHM simulated for all clouds with a total peak brightness temperature greater than $\sim$12$\sigma$, and complete at the same FWHM limit as without a broad component ($\sim$8~km~s$^{-1}$) for clouds greater than 8$\sigma$. 

The completeness is much higher in this case than in the isolated narrow cloud case due to the contribution of flux from the broad component during the binning process ensuring that the narrow cloud was not lost among the noise. In the GASS catalogue of HVCs, we have identified two-component clouds (either featuring a narrow-line component within a broad component or co-located components) with a \textbf{T} in the flag column (Section \ref{results}).

To find any isolated narrow clouds (those without a co-located broad component), we performed a targeted search for narrow clouds on the unbinned GASS~II data, with a peak cutoff of 6$\sigma$ (0.33~K) and a threshold of 2$\sigma$ (0.11~K). This resulted in almost 3500 candidate clouds. We crossmatched these sources with those found by our main source finding method in order to remove any that are already accounted for and removed any with spectral extents greater than 15 channels (the minimum width after binning), which reduced the total number to 532 candidate narrow clouds. The distribution of all of these revealed a high concentration of artifacts at both positive and negative velocities near the southern Galactic pole and along the Galactic plane, as well as some associated with small regions of incompletely flagged RFI. We removed these sources from the candidate list, leaving 54 candidates to be investigated. 

Almost all of these candidates appear at low deviation velocity, with many showing an artificial narrowness due to their proximity to the masked data. Only 14 candidates were kept, which were those showing a well-defined maximum in their spectrum indicating that most of the source is seen. Of the rest, 31 were removed due to their spectra being cutoff by the deviation velocity mask and 9 were removed as noise. We added the 14 sources to our final catalogue, with a flag of \textbf{N} identifying them as separately found narrow-line sources. While this procedure was necessary to ensure our main source finding had not missed a significant population of narrow-line clouds due to the effect of binning the data, we conclude that there were very few narrow clouds missed.
\begin{sidewaystable}
\caption{Excerpt from the GASS HVC catalogue.}
\begin{footnotesize}
\begin{tabular}{l@{~~~}c@{~~~}c@{~~~}r@{~~~}r@{~~~}r@{~~}c@{~~}c@{~~}r@{~~}r@{~~~}r@{~~~}r@{~~~}c@{~~~}c@{~~~}c@{~~~}}
\tableline
\tableline
(1)               & (2)              & (3)              & (4)              & (5)             &  (6)             & (7)         & (8)       & (9)     & (10)   & (11) & (12) & (13) & (14) & (15) \\ 
Name ($l$ \& $b$ \& $v_{\rm LSR}$) & $\alpha$ (J2000) & $\delta$ (J2000)  & $v_{\rm LSR}\pm \Delta$   & $v_{\rm GSR}$           & $v_{\rm dev}$  & ${\rm FWHM}\pm \Delta$ & log$_{10}N_{\rm H}\pm \Delta$ &  $T_{\rm b,fit}$    & Area & $\Delta$x & $\Delta$y & Flags & HIPASS ID & WW91 ID\\ 
(deg \& deg \& km s$^{-1}$) & (${h:m:s}$)      & ($d:m:s$)   & (km s$^{-1}$) &   (km s$^{-1}$)      & (km s$^{-1}$)    & (km s$^{-1}$)  & (cm$^{-2}$) &  (K)  &  (deg$^2$) & (deg) & (deg) & &     \\ 
 \tableline
GHVC G000.1$-$07.3+278 & 18:15:37.92 & -32:26:30 & 278.9$\pm$3.4 & 279.4 & 226.4 & 13.4$\pm$6.7 & 18.56$\pm$18.12 & 0.25 & 0.2 &0.7 &0.5 & - &- &-\\ 
GHVC G000.2$-$11.4$-$239 & 18:33:34.96 & -34:06:55 & -239.8$\pm$4.8 & -238.7 & -192.3 & 20.4$\pm$9.6 & 18.62$\pm$18.31 & 0.20 & 0.1 &0.3 &0.8 & H$_{1}$ & HVC 000.2-11.5-233 & GCN\_GCN \\ 
GHVC G000.6$-$54.7$-$091 & 22:12:32.87 & -40:17:59 & -91.5$\pm$1.8 & -90.1 & -54.0 & 13.6$\pm$3.5 & 18.72$\pm$18.13 & 0.26 & 0.3 &0.9 &0.9 & H$_{2}$ & HVC 000.9-54.8-094 & N\\ 
GHVC G000.5$-$75.7+169 & 23:53:10.86 & -34:00:55 & 169.2$\pm$4.4 & 169.8 & 136.7 & 27.6$\pm$8.7 & 19.08$\pm$18.44 & 0.24 & 0.6 &1.0 &1.0 & H$_{2}$ & HVC 000.5-75.8+173	& -\\ 
GHVC G000.6+21.2$-$103 & 16:31:51.30 & -16:06:53 & -103.3$\pm$2.7 & -101.0 & -60.8 & 25.0$\pm$5.4 & 19.08$\pm$18.39 & 0.32 & 0.5 &0.5 &2.2 & H$_{1}$ & CHVC 000.6+21.3-104	& N\\ 
GHVC G000.9+06.0$-$258 & 17:25:04.98 & -24:52:34 & -258.2$\pm$4.1 & -254.5 & -205.7 & 19.8$\pm$8.2 & 18.79$\pm$18.29 & 0.32 & 0.4 &0.3 &1.5 & H$_{1}$ & HVC 001.1+06.1-264 & GCN\_GCN\\ 
GHVC G001.0$-$66.4$-$094 & 23:10:14.19 & -37:17:08 & -94.7$\pm$4.9 & -93.0 & -62.2 & 20.8$\pm$9.8 & 18.59$\pm$18.31 & 0.21 & 0.2 &1.0 &0.8 & H$_{1}$ & :HVC 001.1-66.4-093 & N\\ 
GHVC G001.2$-$15.4$-$185 & 18:53:02.47 & -34:53:10 & -185.8$\pm$1.0 & -181.2 & -138.3 & 21.4$\pm$2.0 & 19.75$\pm$18.33 & 1.39 & 1.1 &1.0 &1.3 & H$_{1}$ & CHVC 001.2-15.5-186 & GCN\_GCN,N\\ 
GHVC G001.2$-$67.2+122 & 23:14:11.89 & -36:56:46 & 122.5$\pm$2.0 & 124.4 & 90.0 & 16.1$\pm$4.0 & 18.79$\pm$18.20 & 0.31 & 0.4 &1.4 &1.2 & H$_{2}$ & :HVC 001.5-67.3+122 & -\\ 
GHVC G001.4+43.0$-$160 & 15:25:51.75 & -01:37:18 & -160.0$\pm$3.7 & -155.8 & -122.5 & 16.9$\pm$7.5 & 18.61$\pm$18.22 & 0.22 & 0.3 &0.7 &1.0 & H$_{2}$ & HVC 001.6+43.1-161 & N\\ 
\tableline
\end{tabular}\label{table:hvcs}
\end{footnotesize}
\end{sidewaystable}
\begin{sidewaystable}
\caption{Excerpt from the GASS AVC catalogue.}
\begin{footnotesize}
\begin{tabular}{l@{~~~}c@{~~~}c@{~~~}r@{~~~}r@{~~~}r@{~~}c@{~~}c@{~~}r@{~~}r@{~~~}r@{~~~}r@{~~~}c@{~~~}c@{~~~}c@{~~~}}
\tableline
\tableline
(1)               & (2)              & (3)              & (4)              & (5)             &  (6)             & (7)         & (8)       & (9)     & (10)   & (11) & (12) & (13) & (14) & (15) \\ 
Name ($l$ \& $b$ \& $v_{\rm LSR}$) & $\alpha$ (J2000) & $\delta$ (J2000)  & $v_{\rm LSR}\pm \Delta$   & $v_{\rm GSR}$           & $v_{\rm dev}$  & ${\rm FWHM}\pm \Delta$ & log$_{10}N_{\rm H}\pm \Delta$ &  $T_{\rm b,fit}$    & Area & $\Delta$x & $\Delta$y & Flags & HIPASS ID & WW91 ID\\ 
(deg \& deg \& km s$^{-1}$) & (${h:m:s}$)      & ($d:m:s$)   & (km s$^{-1}$) &   (km s$^{-1}$)      & (km s$^{-1}$)    & (km s$^{-1}$)  & (cm$^{-2}$) &  (K)  &  (deg$^2$) & (deg) & (deg) & &     \\ 
 \tableline
GAVC G000.2$-$00.4$-$109 & 17:48:01.89 & -28:54:26 & -109.4$\pm$1.0 & -108.3 & -41.9 & 35.3$\pm$2.0 & 21.15$\pm$18.54 & 16.97 & 145.9 &59.9 &72.0 & C,A,H$_{113}$ & Complex L & N\\ 
GAVC G000.4$-$38.5$-$080 & 20:46:51.33 & -41:09:57 & -80.0$\pm$3.7 & -78.7 & -42.5 & 19.8$\pm$7.3 & 18.65$\pm$18.29 & 0.22 & 0.9 &2.0 &2.1 & T,C,A & - & N\\ 
GAVC G000.9$-$07.0$-$102 & 18:16:05.12 & -31:36:28 & -102.1$\pm$5.6 & -98.6 & -49.6 & 19.3$\pm$11.2 & 18.58$\pm$18.28 & 0.22 & 0.2 &0.5 &0.6 & C,A & - & N\\ 
GAVC G001.3$-$16.6$-$084 & 18:58:45.70 & -35:15:34 & -84.3$\pm$2.3 & -79.4 & -36.8 & 13.4$\pm$4.6 & 18.53$\pm$18.12 & 0.23 & 0.5 &1.3 &1.4 & C,A & - & N\\ 
GAVC G001.4$-$13.7$-$093 & 18:46:02.57 & -34:05:13 & -93.1$\pm$4.3 & -87.8 & -45.6 & 15.5$\pm$8.7 & 18.41$\pm$18.19 & 0.16 & 0.3 &1.0 &0.8 & C,A & - & N\\ 
GAVC G001.4$-$81.5+081 & 00:17:38.69 & -31:17:59 & 81.3$\pm$6.3 & 82.1 & 48.8 & 27.2$\pm$12.6 & 18.76$\pm$18.43 & 0.21 & 0.2 &1.4 &0.7 & C,A & - & -\\ 
GAVC G001.5$-$21.1+075 & 19:20:03.25 & -36:40:12 & 75.3$\pm$1.0 & 80.8 & 32.8 & 6.3$\pm$2.0 & 18.93$\pm$17.79 & 0.56 & 1.7 &2.1 &2.1 & T,C,A & - & -\\ 
GAVC G001.6+32.8$-$073 & 15:57:21.50 & -08:02:41 & -73.7$\pm$1.0 & -68.2 & -36.2 & 11.0$\pm$2.0 & 19.11$\pm$18.04 & 0.67 & 17.2 &9.0 &14.2 & C,A  & - & N\\ 
GAVC G001.8$-$04.1+108 & 18:06:20.81 & -29:26:15 & 108.0$\pm$4.6 & 115.0 & 45.5 & 15.9$\pm$9.2 & 18.40$\pm$18.20 & 0.26 & 0.2 &0.5 &0.7 & C,A & - & -\\ 
GAVC G002.1$-$82.5+084 & 00:21:42.91 & -30:44:22 & 84.3$\pm$4.6 & 85.4 & 51.8 & 22.3$\pm$9.2 & 18.75$\pm$18.34 & 0.22 & 0.3 &1.0 &0.9 & H$_{1}$,A & HVC 000.4-82.5+109 & - \\ 
\tableline
\end{tabular}\label{table:avcs}
\end{footnotesize}
\end{sidewaystable}

\begin{figure*}
  \centering
  \includegraphics[width=0.5\textwidth, angle=-90, trim=0 0 0 0]{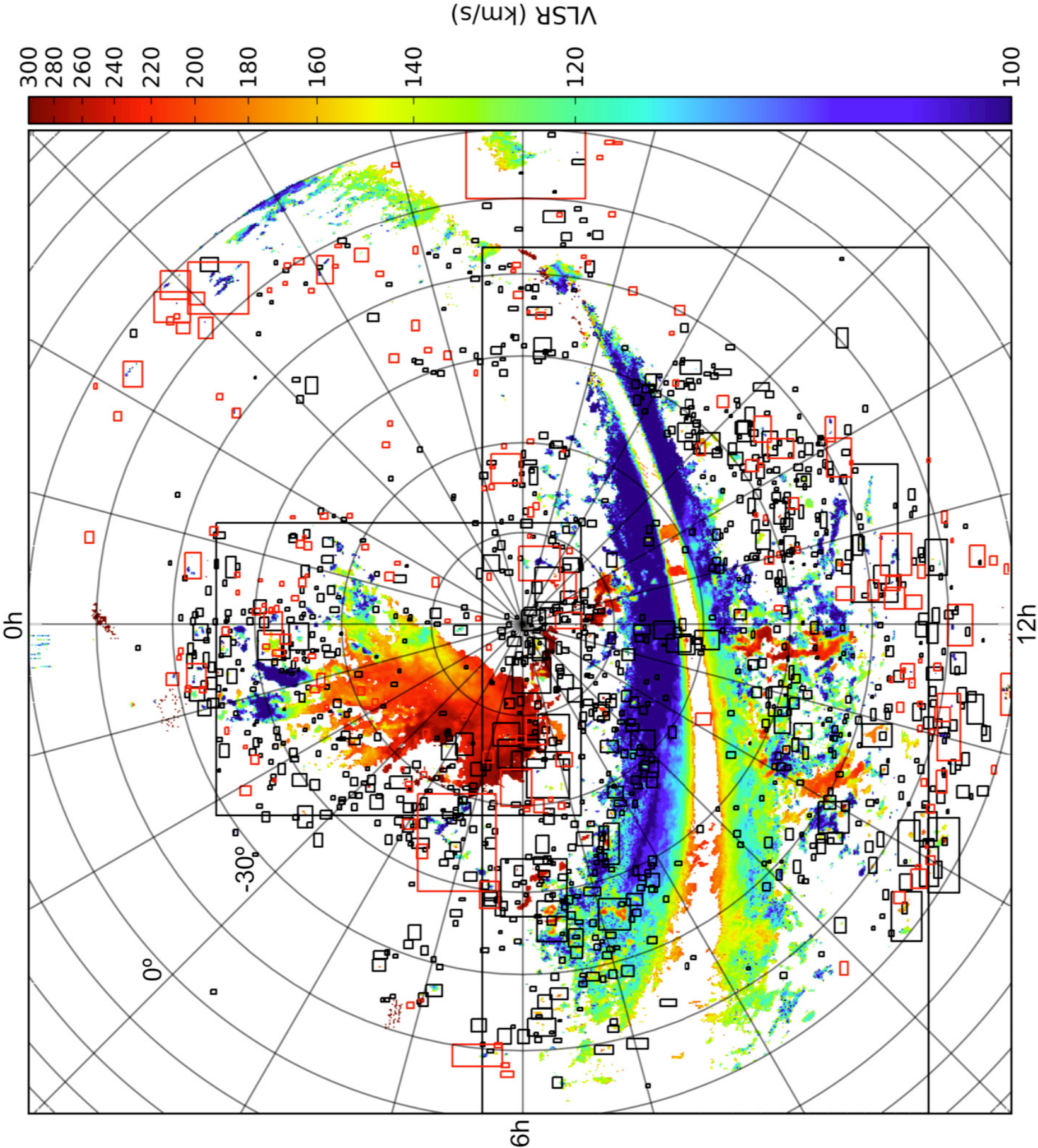}
  \includegraphics[width=0.5\textwidth, angle=-90, trim=0 0 0 0]{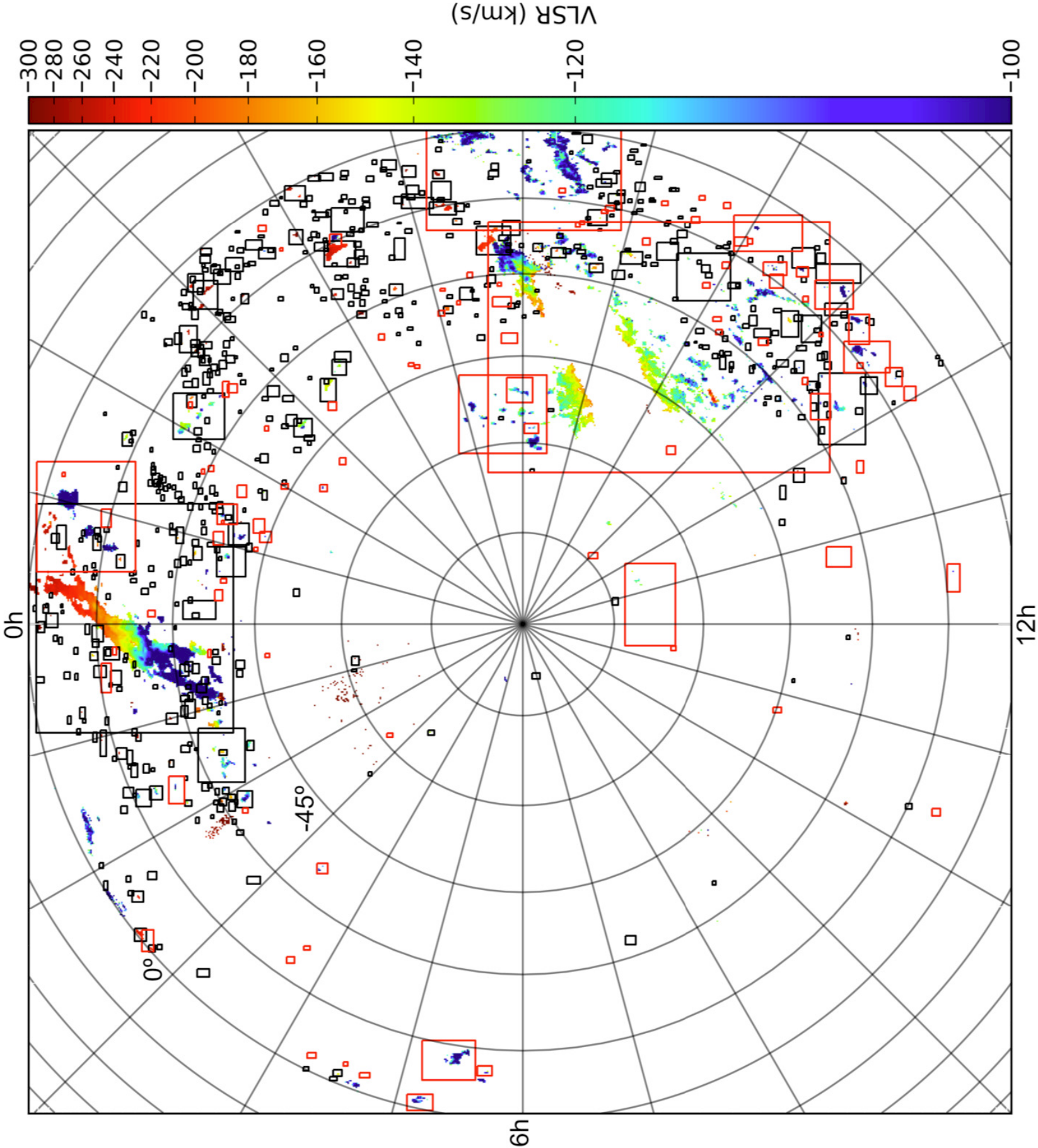}
  \caption{\small The distribution of GASS HVCs and AVCs on the sky, shown overlaid on positive and negative intensity-weighted mean LSR velocity (1st moment) maps (scaled with an arcsinh function).  The projection shown is a zenith-equal-area (ZEA) projection with the south celestial pole at the centre, and each concentric ring moving outwards from the pole represents an increase of 15$^\circ$ in declination. An RA of 0~hours occurs at the 12 o'clock position on the diagram, increasing by 1~hour with each line anticlockwise. We see a strong concentration of clouds near the Magellanic stream, and at latitudes close to the Galactic plane. All emission with $| v_{dev} | <$~30~km~s$^{-1}$, declination $>$ 0$^\circ$, $| v_{LSR} | < 50$~km~s$^{-1}$ or $| v_{LSR} | > 468$~km~s$^{-1}$ is masked out. The colourbar is set to the range 100-300~km~s$^{-1}$ in order to show velocity structure. The two large positive-velocity sources which appear to be unmasked Galactic plane emission contain clouds which deviate from Galactic rotation, and hence were included in our catalogue. HVCs are shown as black boxes, and AVCs are shown as red boxes.}
  \label{fig:dist}
\end{figure*}


\section{The GASS HVC catalogue}\label{results}
Our final HVC catalogue contains a total of 1693 HVCs, including 1111 positive velocity HVCs and 582 negative velocity HVCs. We also include in a separate table a catalogue of all anomalous velocity clouds (AVCs) that do not meet the standard HVC criteria (peak deviation velocity $>$ 50~km~s$^{-1}$ \citep{Wakker:1991p22252} and LSR velocity $>$ 90~km~s$^{-1}$). A cloud only needs to fail one of these velocity criteria to be classed as an AVC. There are a total of 295 AVCs, including 197 positive velocity clouds and 98 negative velocity clouds. We note that these AVCs are likely comprised predominantly of intermediate velocity clouds (IVCs), but as many of these clouds skirt the tenuous boundary between IVC and HVC, we designate them as AVC pending further investigation. The distribution of all positive and negative velocity clouds overlaid on their respective intensity-weighted mean LSR velocity maps is given in Figure \ref{fig:dist}. 

Although the majority of clouds were identified using the binned data, all measurements of observable and derived physical properties were performed on the unbinned but deviation-velocity masked data. Clouds located close to the deviation-velocity cutoff were included only if their spectrum showed a well-defined maximum (indicating the peak of the source), hence distinguishing at least part of their emission from Galactic rotation. This affects 231 clouds, or $\sim$12\% of the total population. 

An excerpt from the GASS HVC catalogue is given in Table \ref{table:hvcs} and an excerpt from the GASS AVC catalogue is given in Table \ref{table:avcs}, with the complete tables available for download as supplementary material in the online-only version of this paper. Below, we give detail on the derivation of each column of Tables \ref{table:hvcs} and \ref{table:avcs}. 

\begin{description}
\item[Column 1] ~\\
The full identifying name of each cloud is given in Column (1). Each cloud is given the identifier GHVC (GASS High Velocity Cloud) or GAVC (GASS Anomalous Velocity Cloud) followed by its truncated Galactic coordinates (based on the position and fitted LSR (Local Standard of Rest) velocity of the best-fit spectrum, see descriptor for Column 2-3).

\item[Column 2-3] ~\\
The position of the peak in equatorial coordinates is given in Column (2) and Column (3). The peak position is determined by comparing the least-squares fitted Gaussian through the location of the peak brightness temperature across the source with the least-squares fitted Gaussian of the peak column density. These peaks are not necessarily the same position as the peak brightness temperature can be skewed positively by the presence of noise, and the peak column density is integrated over the width of the source and so may ignore a true peak that occurs at a position of narrower spectral width. The highest fitted peak is chosen to represent the true peak of the source, to account for any bias due to the influence of noise. Figure \ref{fig:error} demonstrates our means of selecting the best-fit spectrum, which is done based on which fitted peak has a greater height as well as requiring a realistic FWHM. We do not use the column density peak as default in order to avoid biasing against brighter but narrow spectra (which give a lower column density than a broader component). The error in these positions is taken to be the positional accuracy of the GASS data, which is equivalent to the beam of GASS ($16'$). In all cases, the area of the cloud is significantly larger than any error in the position of the peak. The increase in positional error with respect to the signal-to-noise should be noted for all sources with fitted peaks $<$ 0.2~K. 

\item[Column 4] ~\\ 
The same best-fit peak spectrum is used to determine the peak LSR velocity, which is given in Column (4). The uncertainty in the peak velocity is the theoretical estimate of \citet{Taylor:1999p23578}, which is given by
\begin{equation}
\frac{\sigma\textrm{FWHM}}{2{T_{pk}}}.
\end{equation} 

\begin{figure*}
  \centering
  \includegraphics[width=0.45\textwidth, angle=90, trim=0 0 0 0]{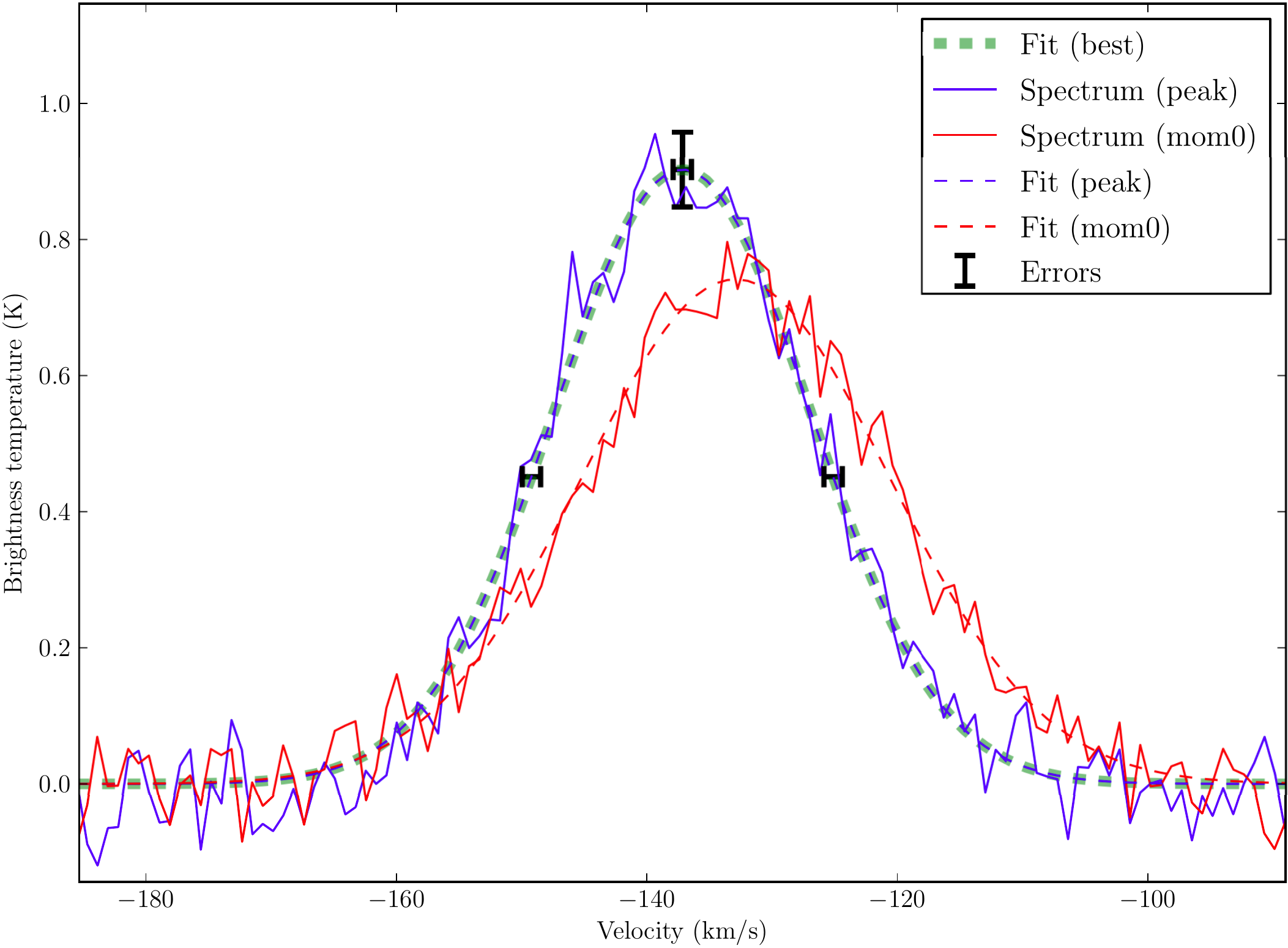}
      \includegraphics[width=0.45\textwidth, angle=90, trim=0 0 0 0]{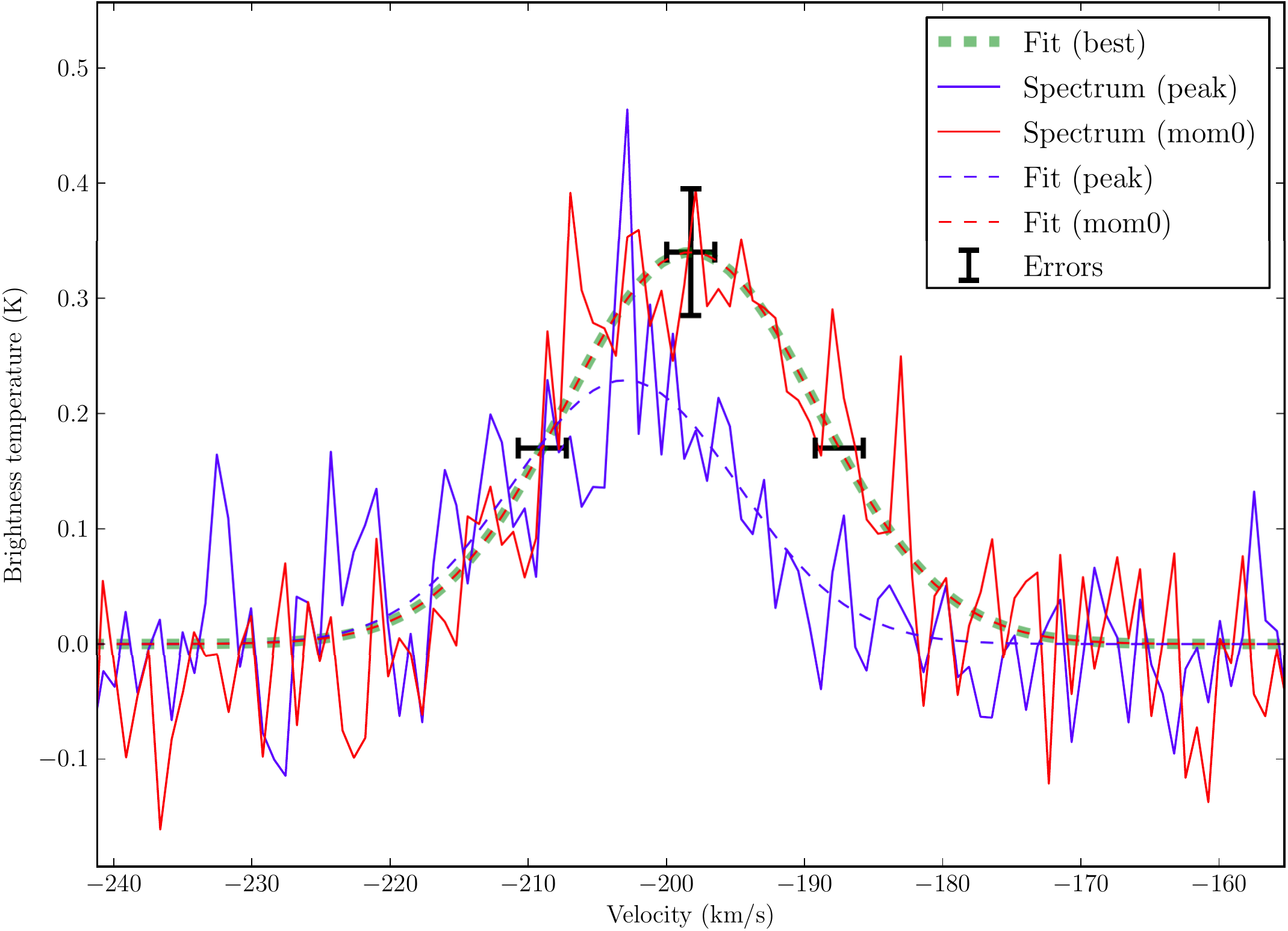}
  \caption{\small Examples of the theoretical errors applied to two separate sources. The top figure shows a high signal-to-noise source, while the bottom figure shows a fairly low signal-to-noise source. The blue curve shows the spectrum obtained at the position of the peak brightness temperature value of the source, while the red curve shows the spectrum at the position of the 0th moment peak (integrated peak). The blue and red dashed lines show the Gaussian fit to the peak and 0th moment spectra respectively. In the top figure, the peak spectrum is selected as the best-fit (based on the lower signal in the 0th moment spectrum), while the 0th moment spectrum is selected as the best-fit in the bottom figure due to the higher fitted peak (which averages out the effect of noise on top of each spectrum). The error bars in both cases represent the theoretical errors on the peak value, the peak position and the FWHM of the source. We see in each case that the theoretical errors are good approximations of the error on our fitted values, based on comparing them to the original spectrum.}
  \label{fig:error}
\end{figure*}

\begin{table*}
\centering
\begin{tabular}{l@{~~~}c@{~~~}c@{~~~}c@{~~~}c@{~~~}c@{~~~}c@{~~~}}
\tableline
\tableline
Property & Mean$_{HVCs}$ & Mean$_{AVCs}$ & Median$_{HVCs}$ & Median$_{AVCs}$ & $\sigma_{HVCs}$ & $\sigma_{AVCs}$\\
\tableline
$|v_{LSR}|$ (km~s$^{-1}$)&	182					&	83					&	167					&	82					&	71					&	13	\\
$|v_{dev}|$ (km~s$^{-1}$)	&	139					&	42					&	123					&	41					&	70					&	6	\\
$|v_{GSR|}$ (km~s$^{-1}$)&	98					&	90					&	78					&	78					&	86					&	61	\\
FWHM (km~s$^{-1}$)	&	19					&	16					&	19					&	16					&	6					&	6	\\
T$_{b,fit}$ (K)			&	0.37					&	0.46					&	0.15					&	0.19					&	3.5					&	1.4	\\
N$_H$ (cm$^{-2}$)		&	2$\times$10$^{19}$		&	2$\times$10$^{19}$		&	5$\times$10$^{18}$		&	5$\times$10$^{18}$		&	3$\times$10$^{20}$		&	2$\times$10$^{20}$	\\
Area (deg$^2$)		&	2.5					&	4.2					&	0.3					&	0.7					&	57					&	20	\\
\tableline
\end{tabular}
\caption{Statistical properties of the GASS HVC catalogue. The mean, median and standard deviation ($\sigma$) are shown for both the HVC and AVC populations.}\label{table:properties}~\\
\end{table*}

\item[Column 5-6] ~\\ 
The uncertainty in LSR velocity also applies to the error in the velocity in GSR (Galactic Standard of Rest) coordinates given in Column (5) and to the deviation velocity given in Column (6). The GSR velocity is calculated using the standard V$_{\rm GSR}$~=~$220 \cos b \sin l +$ V$_{\rm LSR}$. We estimate the deviation velocity based on the same model of Galactic rotation used to mask low deviation velocities in Section \ref{method}. 

\item[Column 7] ~\\ 
The FWHM is measured using a least-squares Gaussian fitting routine, and is given in Column (7). As this was performed using a single-component fit for all sources to ensure consistency, the fitted FWHM should be treated with caution for all sources flagged in Column (11) as having two or more spectral components. The theoretical error of \citet{Taylor:1999p23578} is given for the FWHM, and is defined as
\begin{equation}
\frac{\sigma\textrm{FWHM}}{{T_{pk}}}.
\end{equation} 
We give an example of the peak and FWHM errors applied to both a high signal-to-noise source and a low signal-to-noise source in Figure \ref{fig:error}. 

\item[Column 8] ~\\ 
The peak column density is given in Column (8). This is determined on the basis of measuring the peak integrated intensity from the 0th moment map, and using the standard conversion factor of $1.8 \times 10^{18}$ to convert from K~km~s$^{-1}$ to cm$^{-2}$ of column density (assuming negligible optical depth which may not be true in all cases). The error on this value is assumed to be the nominal GASS noise of 57~mK multiplied by the FWHM of each source. 

\begin{figure*}
  \centering 
    \includegraphics[width=0.35\textwidth, angle=-90, trim=0 0 0 0]{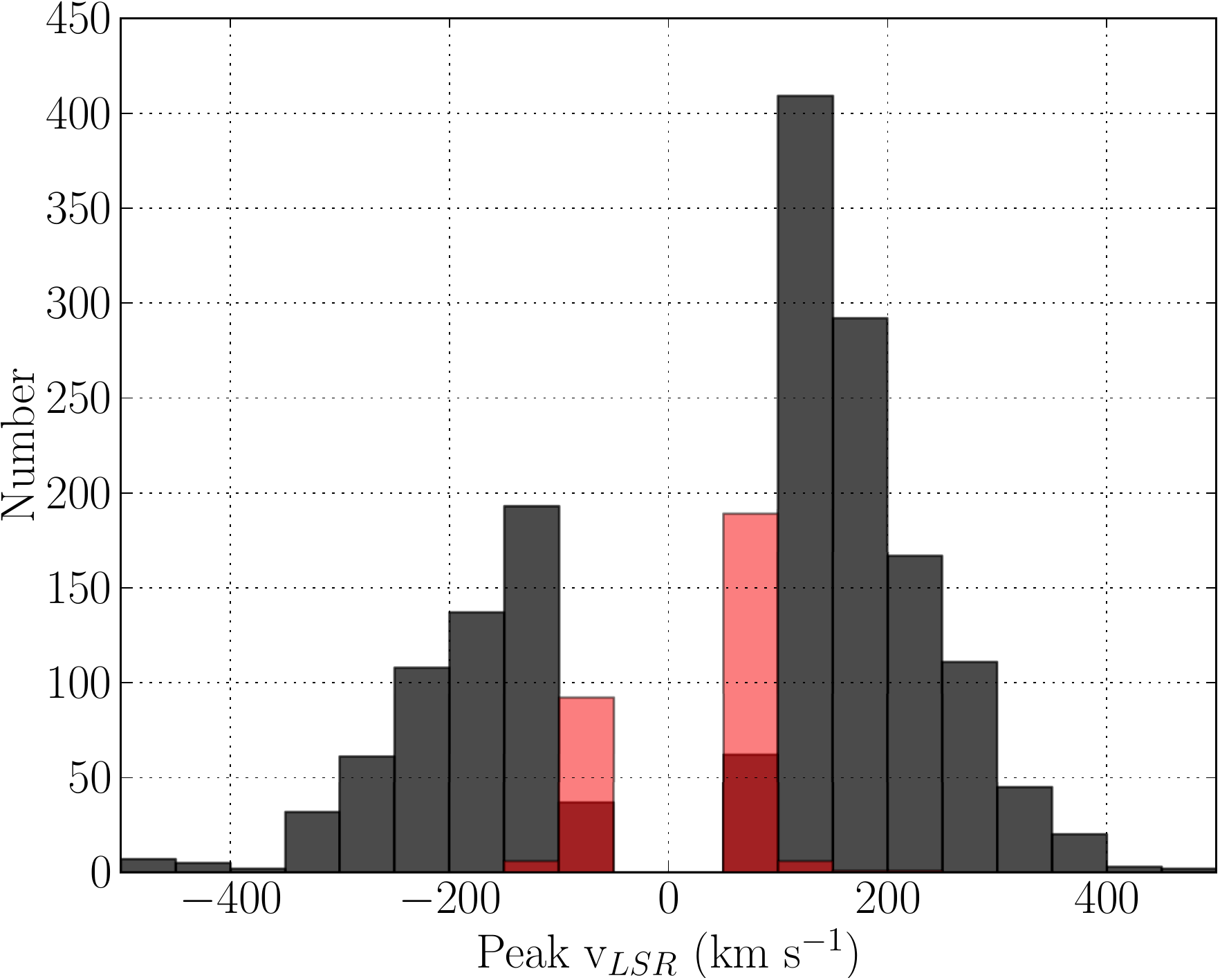}
  \includegraphics[width=0.35\textwidth, angle=-90, trim=0 0 0 0]{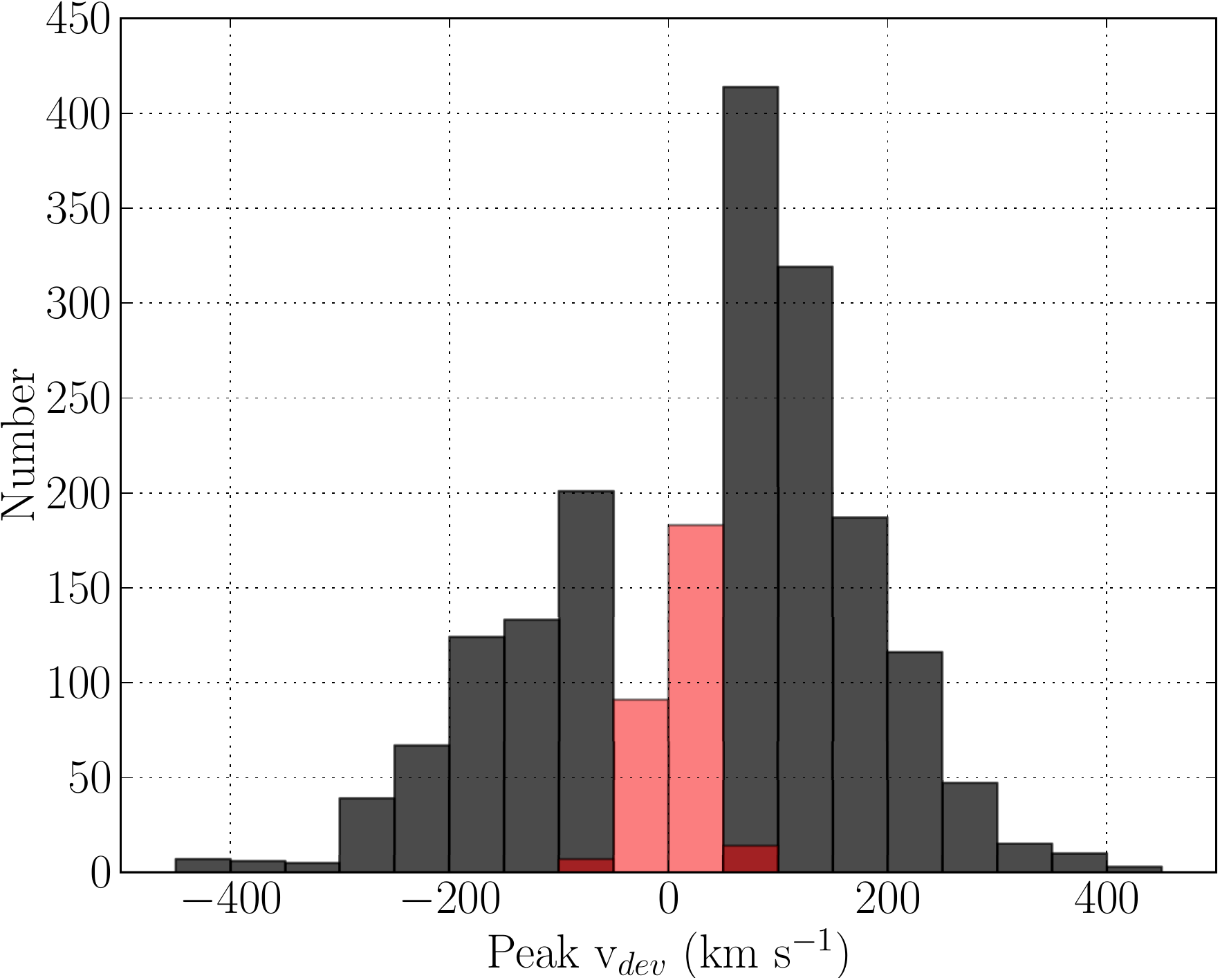}
    \includegraphics[width=0.35\textwidth, angle=-90, trim=0 0 0 0]{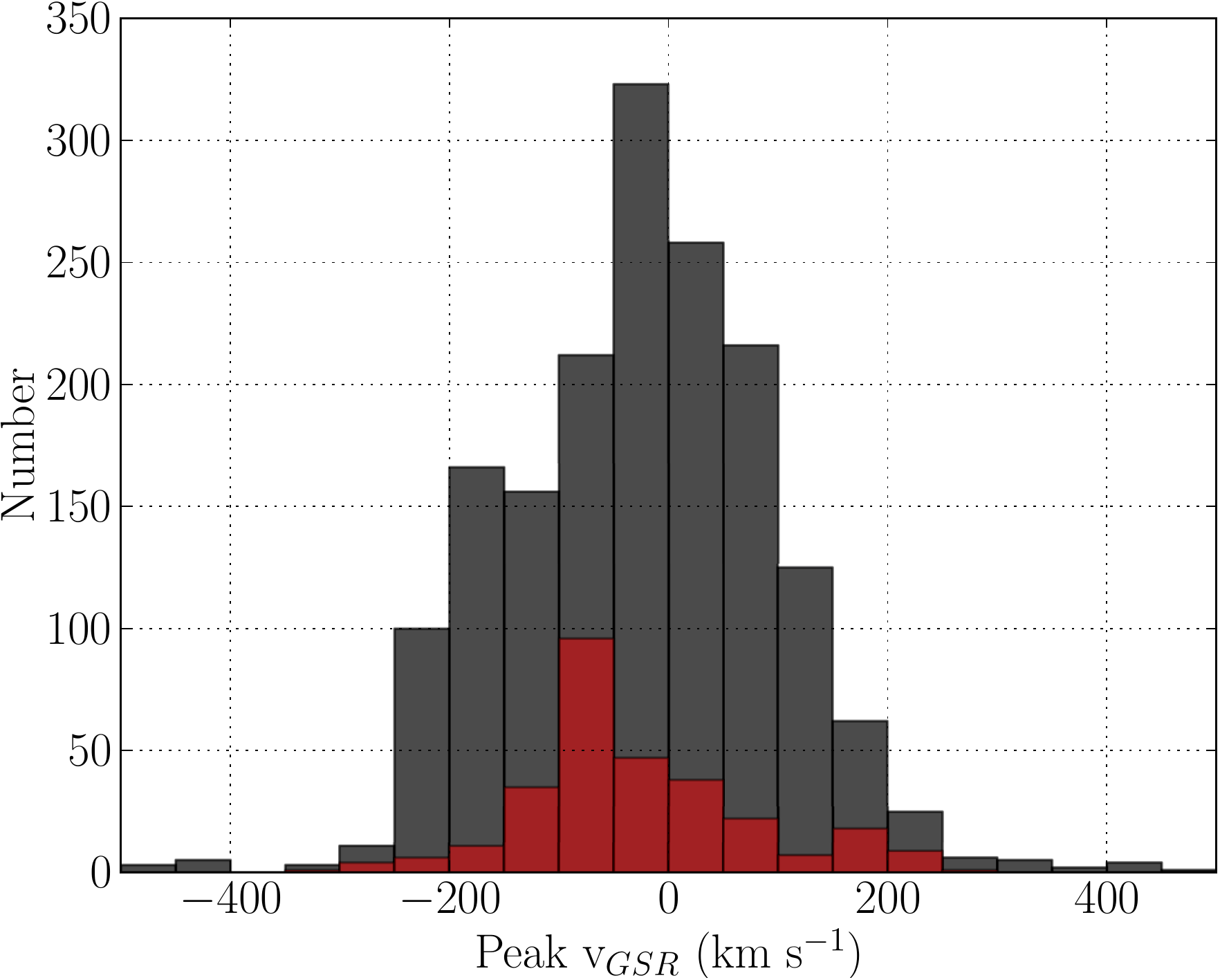}
      \includegraphics[width=0.35\textwidth, angle=-90, trim=0 0 0 0]{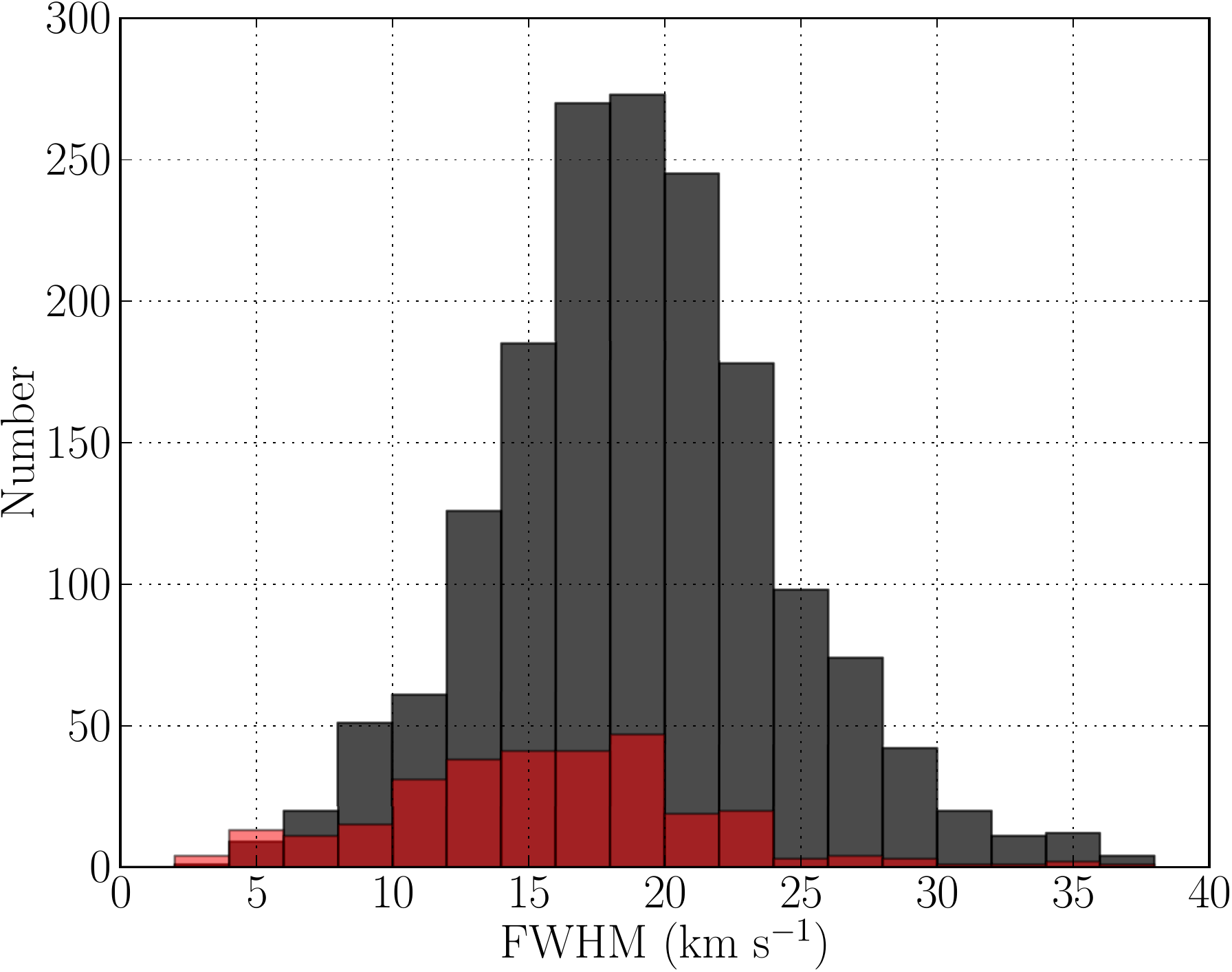}

  \caption{\small Histogram distributions of properties of GASS HVCs and AVCs. In order from top left, across: velocity (Local Standard of Rest), deviation velocity, velocity (Galactic Standard of Rest), and FWHM (single component fit). All velocities are given in relation to the best-fit spectrum (either the peak brightness temperature or the peak column density). In all cases, HVCs are shown in black and AVCs are shown in red.}
  \label{fig:hists}
\end{figure*}

\begin{figure*}
  \centering 
      \includegraphics[width=0.35\textwidth, angle=-90, trim=0 0 0 0]{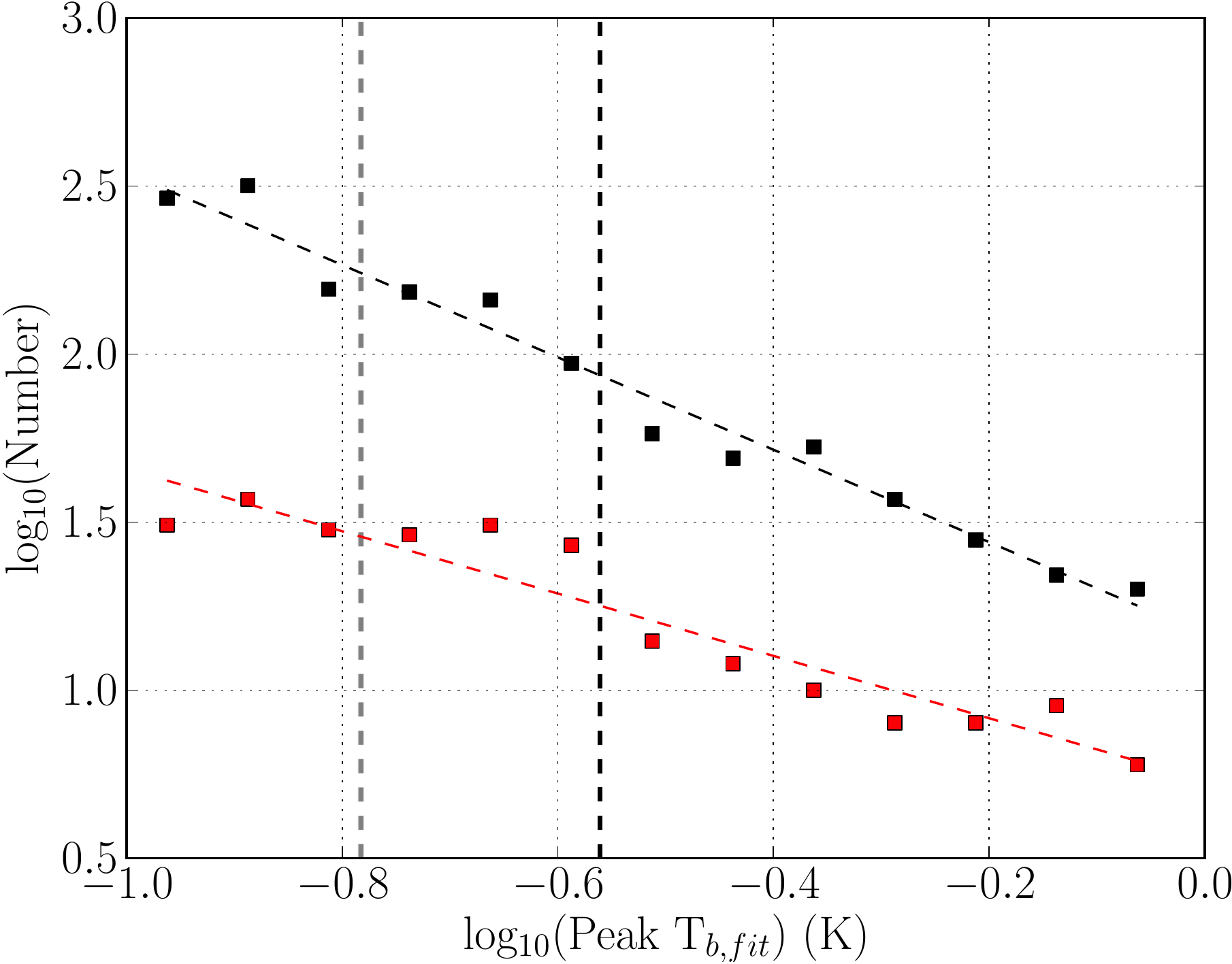}
  \includegraphics[width=0.35\textwidth, angle=-90, trim=0 0 0 0]{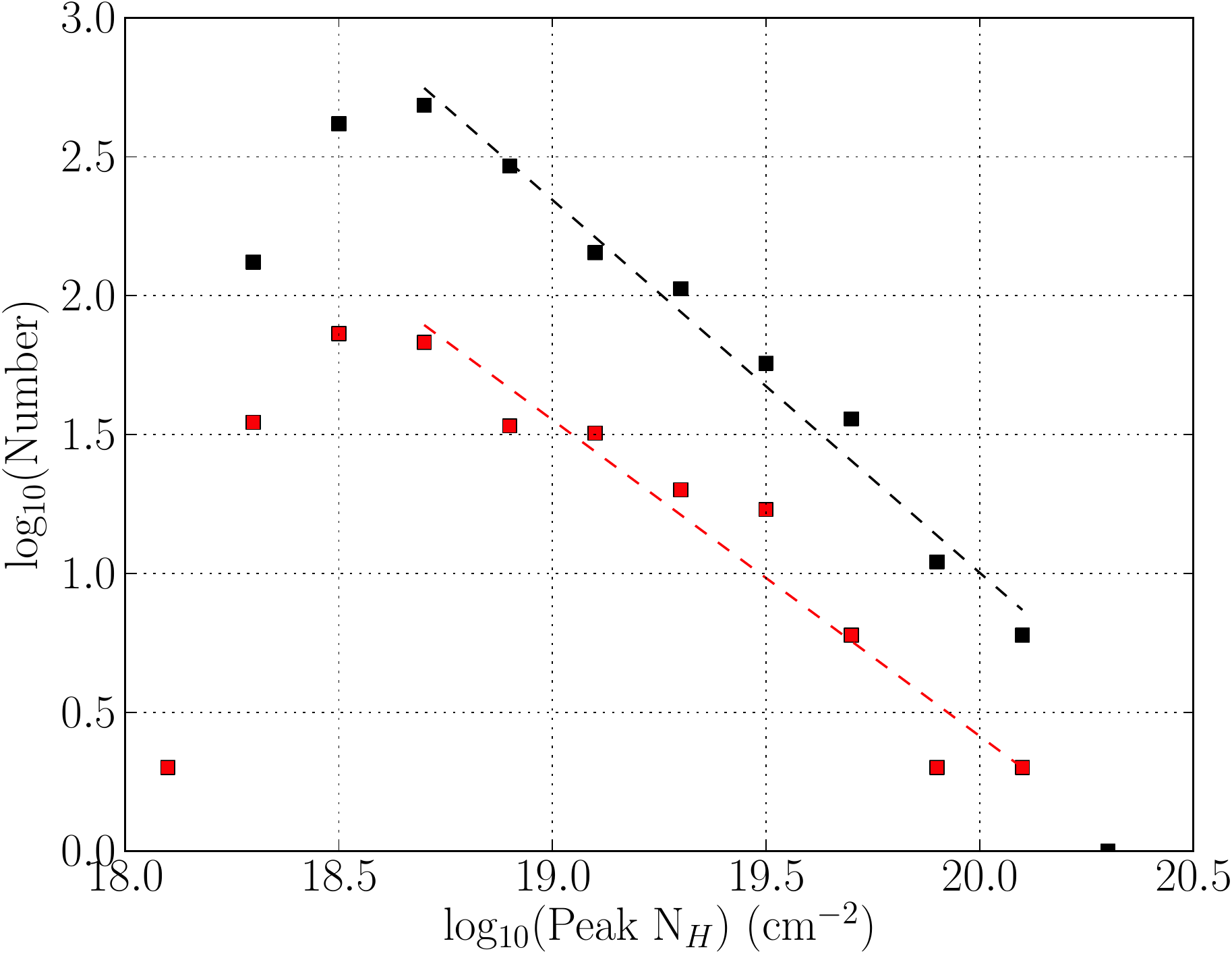}
      \includegraphics[width=0.35\textwidth, angle=-90, trim=0 0 0 0]{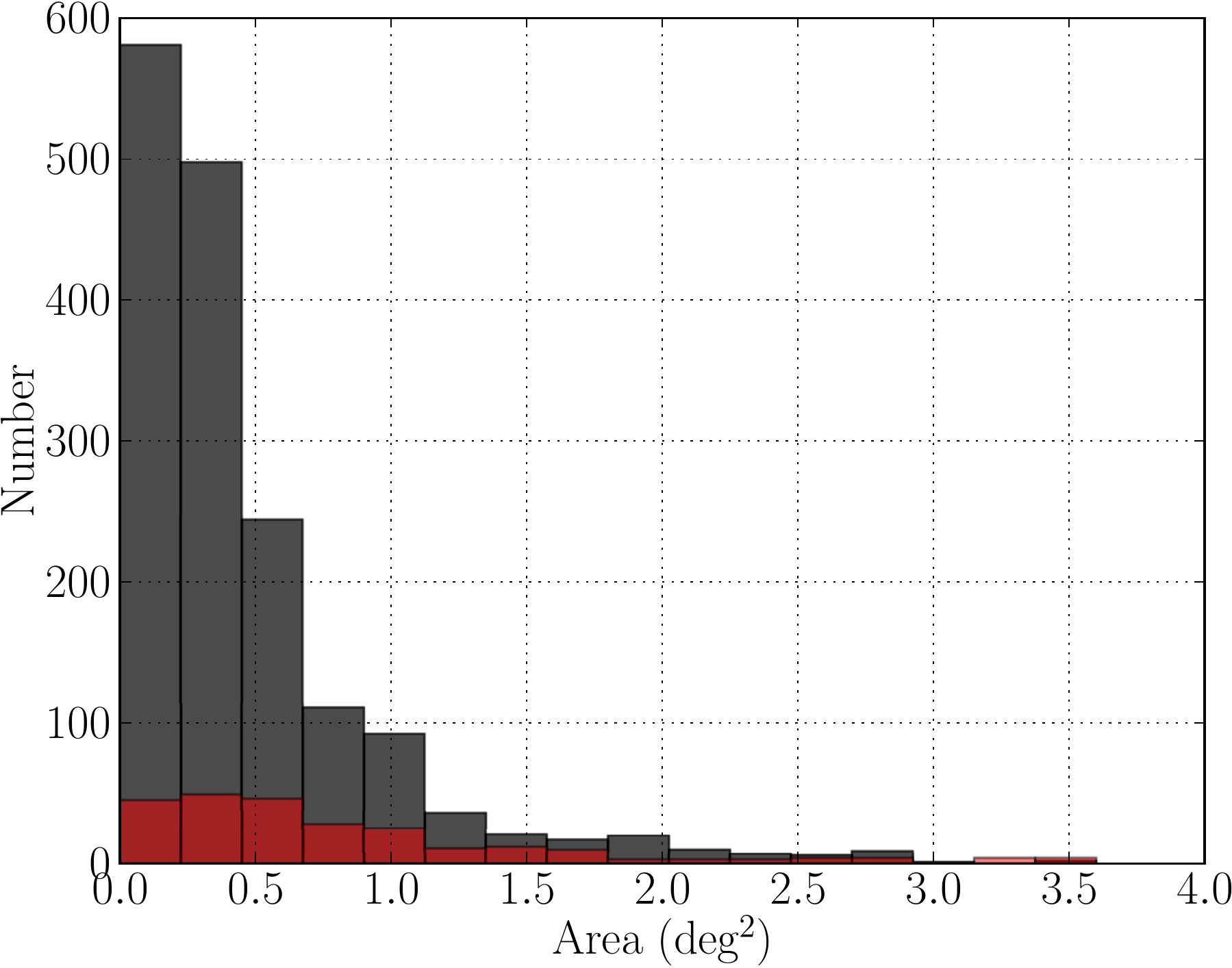}
  \caption{\small Histogram distributions of properties of GASS HVCs and AVCs. In order from top left, across:  brightness temperature (fitted), peak column density, and area. The peak brightness temperature is the fitted value as given in the catalogue, with vertical dashed lines indicating 3$\sigma$ and 5$\sigma$. The peak column density is measured at the peak in the integrated intensity map, and so is not necessarily at the same position as the fitted peak brightness temperature. A linear fit in log-space is given for brightness temperature (HVC slope = $-$1.4, AVC slope = $-$0.9) and column density (HVC slope = $-$1.3, AVC slope = $-$1.1). In all cases, HVCs are shown in black and AVCs are shown in red.}
  \label{fig:hists2}
\end{figure*}

\begin{figure*}
  \centering
  \includegraphics[width=0.5\textwidth, angle=-90, trim=0 0 0 0]{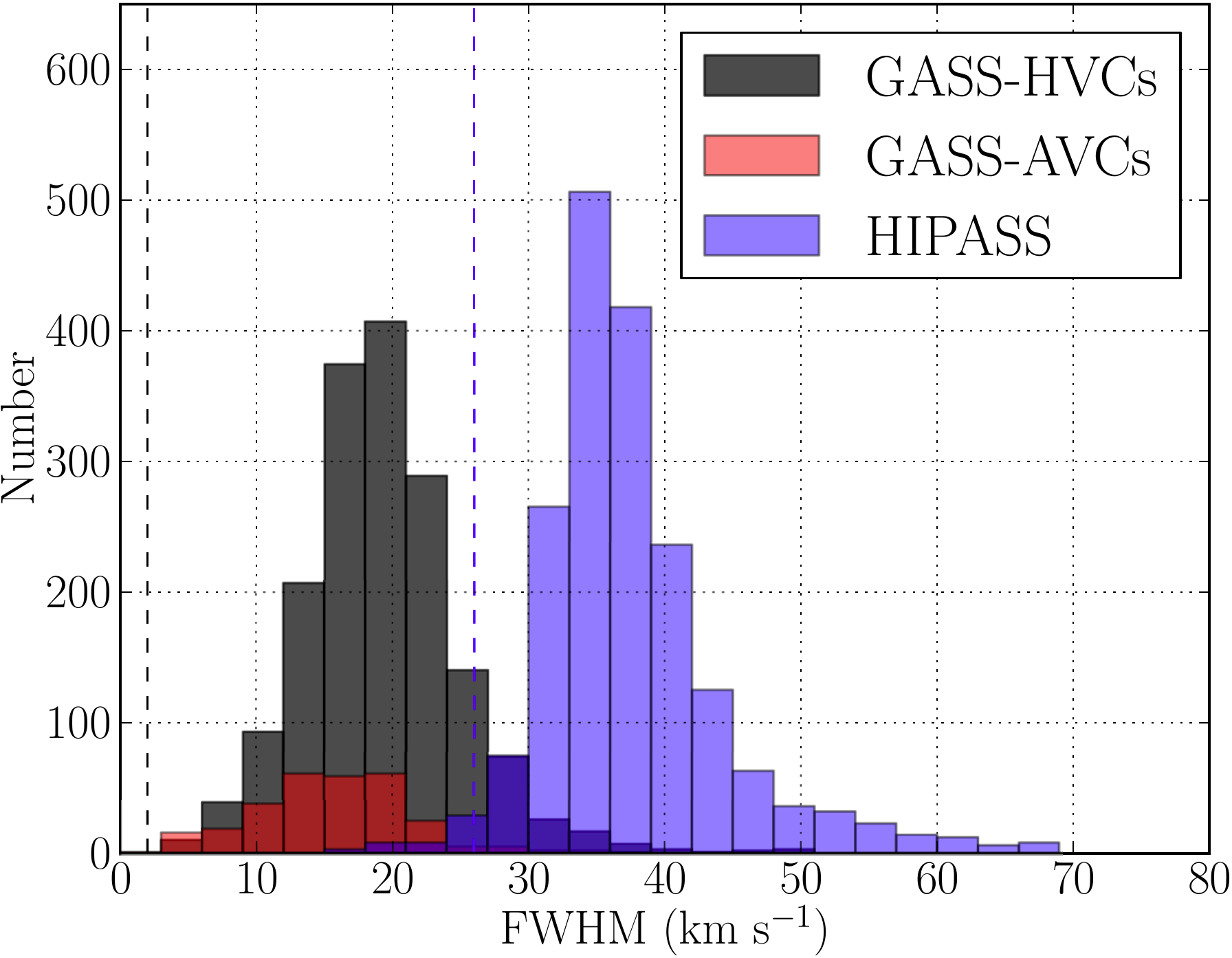}
    \includegraphics[width=0.5\textwidth, angle=-90, trim=0 0 0 0]{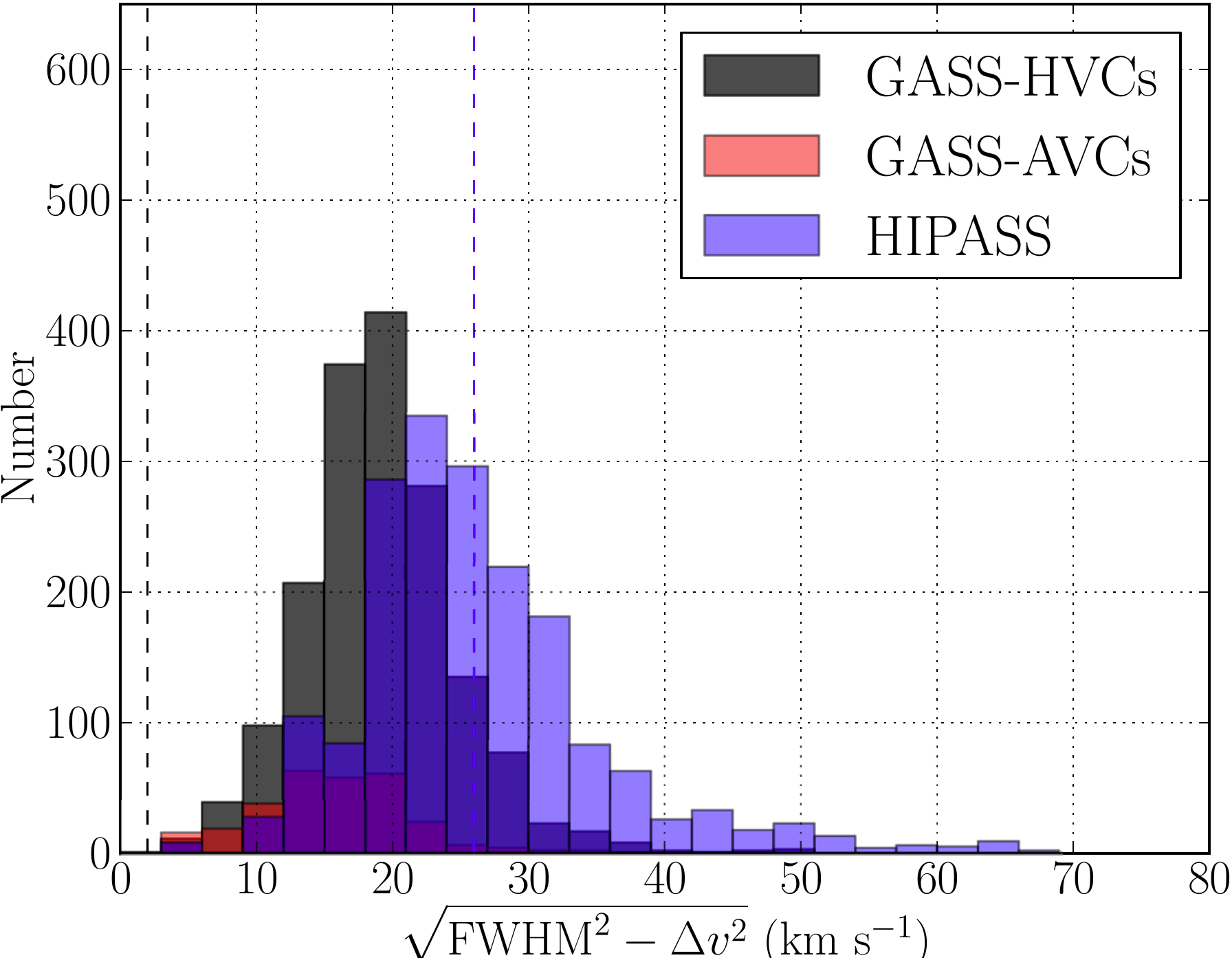}
  \caption{\small The distribution of cloud FWHMs measured through the best-fit spectrum within the GASS HVC catalogue showing HVCs (black) and AVCs (red), compared with the FWHM distribution of the HIPASS HVCs (blue). The respective spectral resolution (incorporating source finding parameters) is shown in the same colour for each survey. The top figure shows the measured FWHMs from both GASS and HIPASS highlighting the effects of differing velocity resolution, while the bottom figure shows the corrected cloud line-widths taking into account the velocity resolution of each respective survey: FWHM$_{observed}$ = FWHM$_{actual}^2 - \Delta v^2$, where $\Delta v$ = 1~km~s$^{-1}$ for GASS and $\Delta v$ = 26.4~km~s$^{-1}$ for HIPASS. We see fairly good agreement between the GASS and HIPASS distributions of cloud FWHM in the case of the corrected line-widths.}
  \label{fig:fwhm}
\end{figure*}

\item[Column 9] ~\\ 
We give the fitted peak brightness temperature of the best-fit spectrum in Column (9) of the table. The fitted peak is given to account for the possible effect of noise, noting that this is based on the assumption of the source as a single-component Gaussian and as such may underestimate the true peak in cases of complex spectra. The error on the peak value is taken to be the RMS of the GASS data, which is in all cases assumed to be 57~mK. 

\item[Column 10] ~\\ 
The area of each source is given in Column (10). This area is determined starting from the rough area of the source on the sky, shown as rectangles in Figure \ref{fig:dist}. We estimate the noise level $\sigma_{0}$ in the 0th moment map as the spectral width of the cube containing the cloud multiplied by 57~mK, and count all pixels above 2$\sigma_0$. The area is the total number of pixels multiplied by the pixel area of 0.08$^\circ \times 0.08^\circ$. It is not practical in our case to perform angular fitting for the majority of sources, as many are either complexes of several clouds or resolved with structure (and hence not well approximated as Gaussians). 

\item[Column 11-12] ~\\ 
The maximum extent of the source above 2$\sigma_0$ is given as $\Delta$x and $\Delta$y in Columns (11) and (12). These dimensions correspond to the $x$-extent and $y$-extent of each source in the original zenith-equal-area (ZEA) coordinate system, and are included to give an idea of the source dimensions. The area given in Column (10) will generally be smaller than the combination of $\Delta$x and $\Delta$y because it only considers the pixels above 2$\sigma_0$ as part of the source.

\item[Column 13] ~\\  
The flags in Column (13) are one or more of \textbf{T}, \textbf{H}, \textbf{C}, \textbf{A}, \textbf{N}. They have the following meaning: A cloud is determined to be better represented by two components if the standard deviation of the single-component residual is $> 10$\% higher than the standard deviation of the two-component residual, and is marked as two-component ({\bf T}). These two-component sources were also checked to make sure they were accurately classified, and removed from the group if their spectrum did not show convincing two component structure. A flag of \textbf{H$_N$} indicates that the region the source occupies contains $N$ HIPASS sources. Sources which sit near the edge of our 30~\kms~deviation velocity cutoff are required to show a well-defined maximum in their spectra, but due to increased uncertainty in their properties (such as FWHM and $N_{\rm H}$) we also flag any partially-cutoff sources with a {\bf C}. If the source does not meet the velocity cutoff criteria of an HVC, it is thus classified as an anomalous velocity cloud (AVC) and is labelled with an {\bf A} as well as its name changing from GHVC to GAVC. The {\bf A} flag is not the same as the {\bf C} flag because it is possible for a source to not meet the HVC velocity criteria and still be significantly away from the deviation velocity mask (and hence not cutoff), however any source with a {\bf C} flag will always have an {\bf A} flag. A flag of \textbf{N} indicates a cloud found in the targeted narrow cloud search rather than in the main source finding process.

\item[Column 14] ~\\ 
For GASS sources containing HIPASS clouds, Column (14) gives the name of the `best match'  in HIPASS. This was done using the results of crossmatching with the HIPASS catalogue of HVCs \citep{Putman:2002p12244} in Section \ref{completeness}. The best match HIPASS cloud indicates that the corresponding GASS cloud is the closest in spatial and spectral coordinates to the specific HIPASS cloud. For the largest GASS sources which contain many HIPASS sources, we list the name of the region (e.g. Magellanic Clouds) rather than the best match cloud. If the flag column indicates a HIPASS source but the HIPASS ID is blank, this means that although the GASS cloud contains the HIPASS source, a different GASS cloud is a closer match. 

\item[Column 15] ~\\ 
Column (15) lists all complexes and populations as defined by \citet{Wakker:1991p22252} (Table 3) that the particular GASS cloud may be associated with based on its Galactic longitude, Galactic latitude and $v_{LSR}$ velocity. If a cloud is listed to be potentially part of several complexes, this is due to overlap in the longitude/latitude/velocity definitions of regions. 

\end{description}

\subsection{Overall properties}
We have summarised the key statistical values for each cloud property in Table \ref{table:properties}, with histograms for properties of the GASS HVCs shown in Figure \ref{fig:hists} and Figure \ref{fig:hists2}. In each histogram plot, we have separated the HVCs (shown in black) from the AVCs (shown in red). The very largest GASS sources (e.g. the two large positive-velocity sources which appear to be unmasked Galactic plane emission seen in Figure \ref{fig:dist}) are a unique population due to their large area on the sky and were not included in the statistics or histograms. There are six of these large regions in total, covering the Magellanic Clouds, Magellanic Stream, Magellanic Leading Arm, Complex L and the Galactic Centre region. 

\begin{figure*}
  \centering
  \includegraphics[width=0.7\textwidth, angle=0, trim=0 0 0 0]{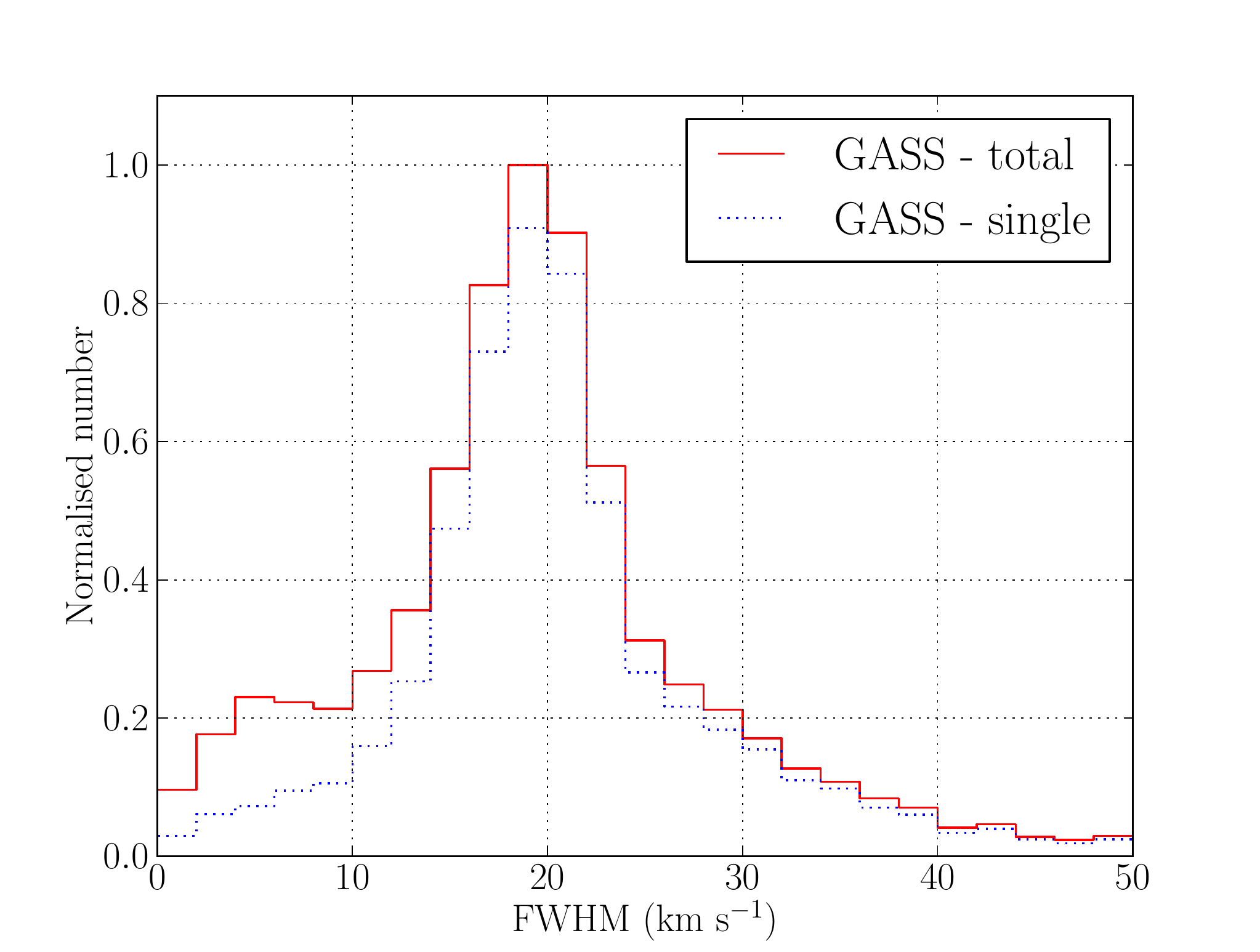}
  \caption{\small The normalised distribution of the measured FWHMs of all spectral components across all GASS HVCs above 5$\sigma$ and $<$ 5 deg$^2$ in angular size. The red line is the distribution of all 13593 spectral components, while the blue line shows the 10767 spectral components whose spectra were identified as single component. We have excluded any spectra with measured velocities of $<$ 90~km~s$^{-1}$. The median for this distribution of 21~km~s$^{-1}$ (FWHM) is close to that of our catalogued FWHM distribution. We see two distinct populations in the total distribution, one centred at $\sim$5~km~s$^{-1}$ and one centred at $\sim$20~km~s$^{-1}$. We also see evidence of structure within the single component distribution, with a narrow concentration at $\sim$20~km~s$^{-1}$ as well as a broader distribution that peaks at a slightly higher velocity width.}
  \label{fig:components}
\end{figure*}

\begin{figure*}
  \centering
  \includegraphics[width=0.35\textwidth, angle=-90, trim=0 0 0 0]{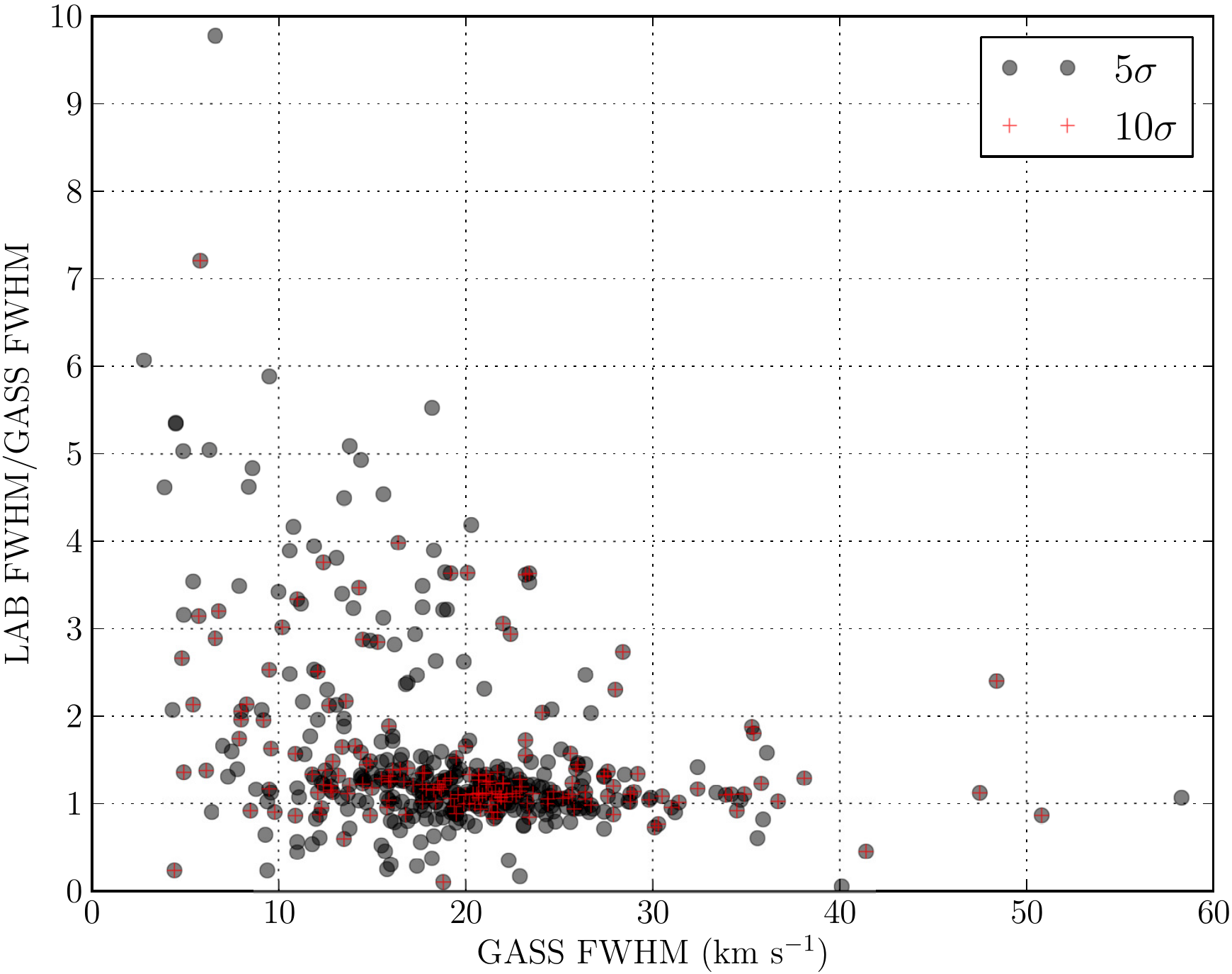}
    \includegraphics[width=0.35\textwidth, angle=-90, trim=0 0 0 0]{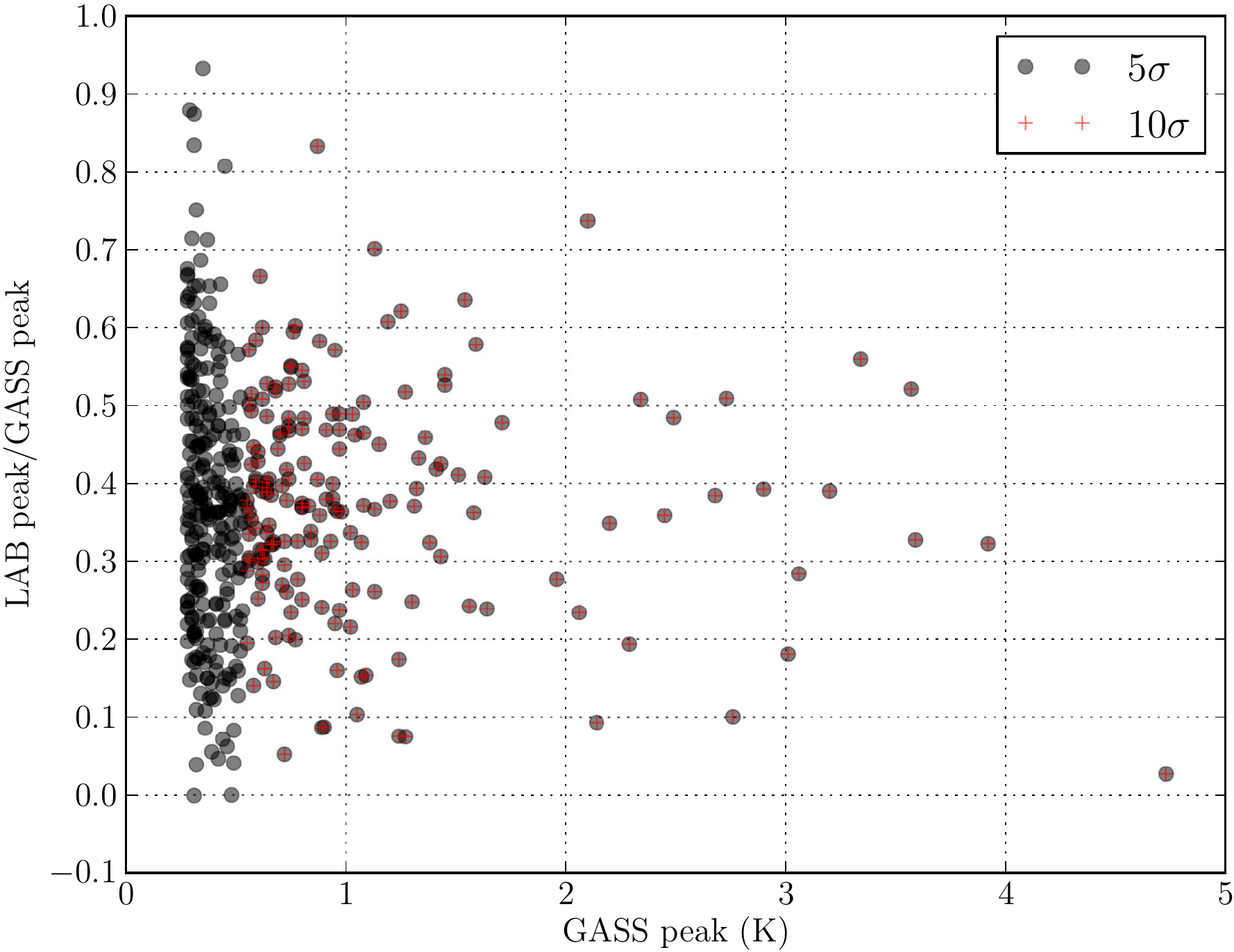}
  \caption{\small Comparative plots showing the difference between the measured FWHMs in GASS and LAB (left) and the difference between the measured peaks in GASS and LAB (right). Equivalence occurs where the y-axis is equal to one. The black circles represent all sources with a peak fitted brightness temperature in GASS of $>$ 5$\sigma$, while the red crosses are the sources $>$ 10$\sigma$ (a subset of the black circles). There is fairly good agreement between GASS and LAB FWHMs, although the distribution is positively shifted indicating a slightly larger FWHM in the LAB data. Because the LAB beam is over 2 times that of GASS, we see that the measured peak in LAB is always less than that of GASS, indicating beam dilution.}
  \label{fig:labcompare}
\end{figure*}

There is no obvious trend in right ascension or declination for either of the populations, although a slightly increased number of sources near right ascension of 0$^\circ$ is likely due to the influence of the Magellanic Stream. There is an evident gap between 100 to 150$^\circ$ in the AVC distribution due to the effect of deviation velocity masking near the Galactic plane (which occupies most of this angle range). The declination distribution tails off at declinations close to 0$^\circ$ because of the increased noise in the GASS data at declinations that were observed at low elevation by the Parkes telescope. This effect is far less exaggerated for the AVCs, which may be a product of the lower number of AVCs as well as the influence of the Magellanic system on the HVC distribution at declinations $\sim$$-$40$^\circ$. Similarly, there are no clear trends in Galactic latitude or longitude due to the limitations of a southern-sky only survey.

All velocity properties are shown in Figure \ref{fig:hists}. The distribution in LSR velocity and deviation velocity is dominated by positive clouds due partly to the influence of the Magellanic Clouds and Stream, and partly because the southern view of the Galaxy is also dominated by gas at positive velocities. The gap in the middle of both velocity histograms is due to the effect of searching only at deviation velocities higher than 30~km s$^{-1}$. The AVC population occupies the low end of both deviation velocity and LSR velocity by definition, although it is worth noting that the AVC population is dominated by sources which have both low deviation velocities \textit{and} low LSR velocities, which is not necessarily to be expected.

We see a preference towards negative velocities with respect to the Galactic Standard of Rest (GSR) in both HVCs and AVCs, although this is evident as an asymmetric distribution for HVCs and as a shifted distribution for AVCs. This bias towards negative velocities is consistent with previous findings in both neutral hydrogen \citep{Stark:1992p23268} and ionised hydrogen \citep{Haffner:2003p20906}. 

\begin{figure*}
  \centering
  \includegraphics[width=0.35\textwidth, angle=-90, trim=0 0 0 0]{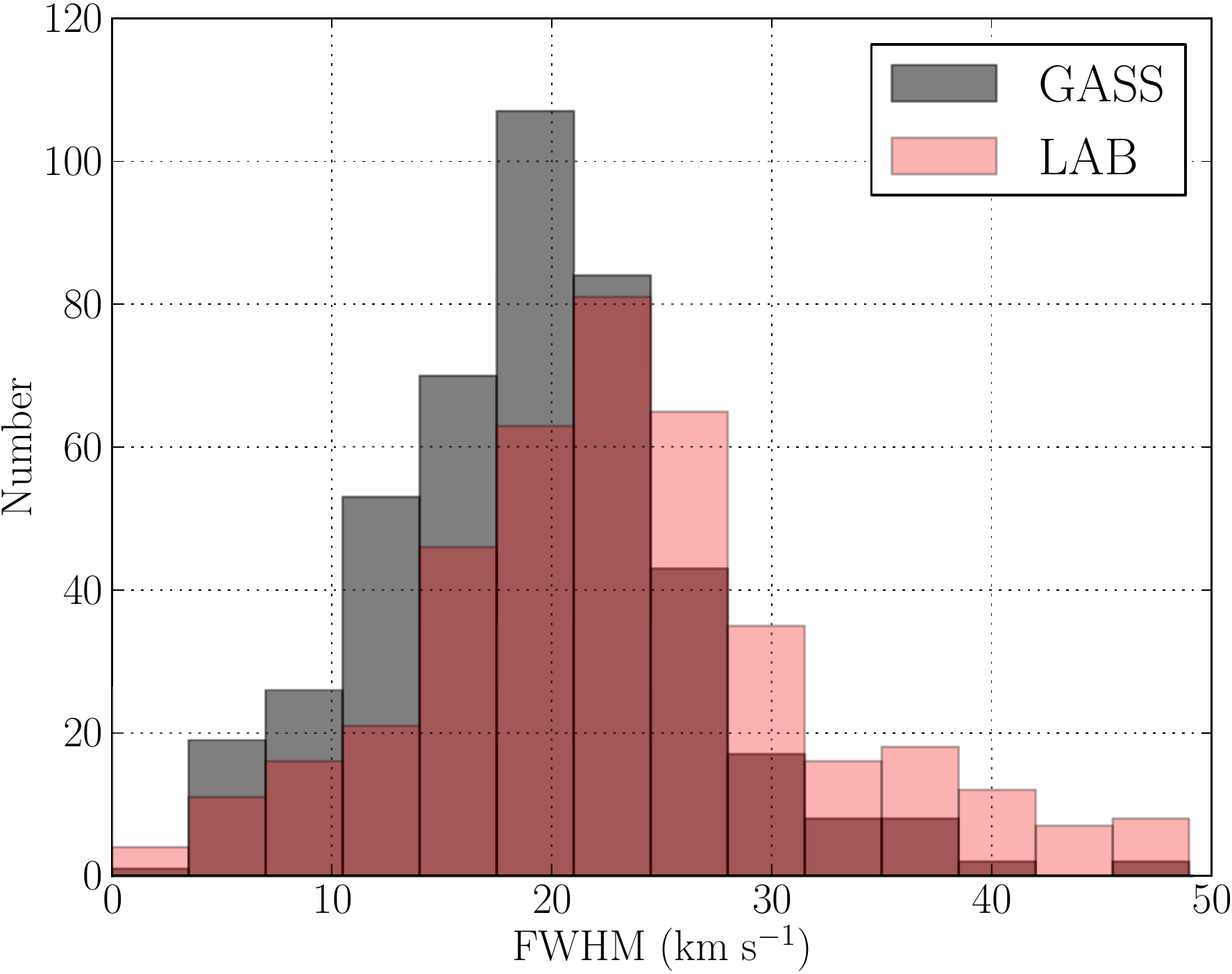}
    \includegraphics[width=0.35\textwidth, angle=-90, trim=0 0 0 0]{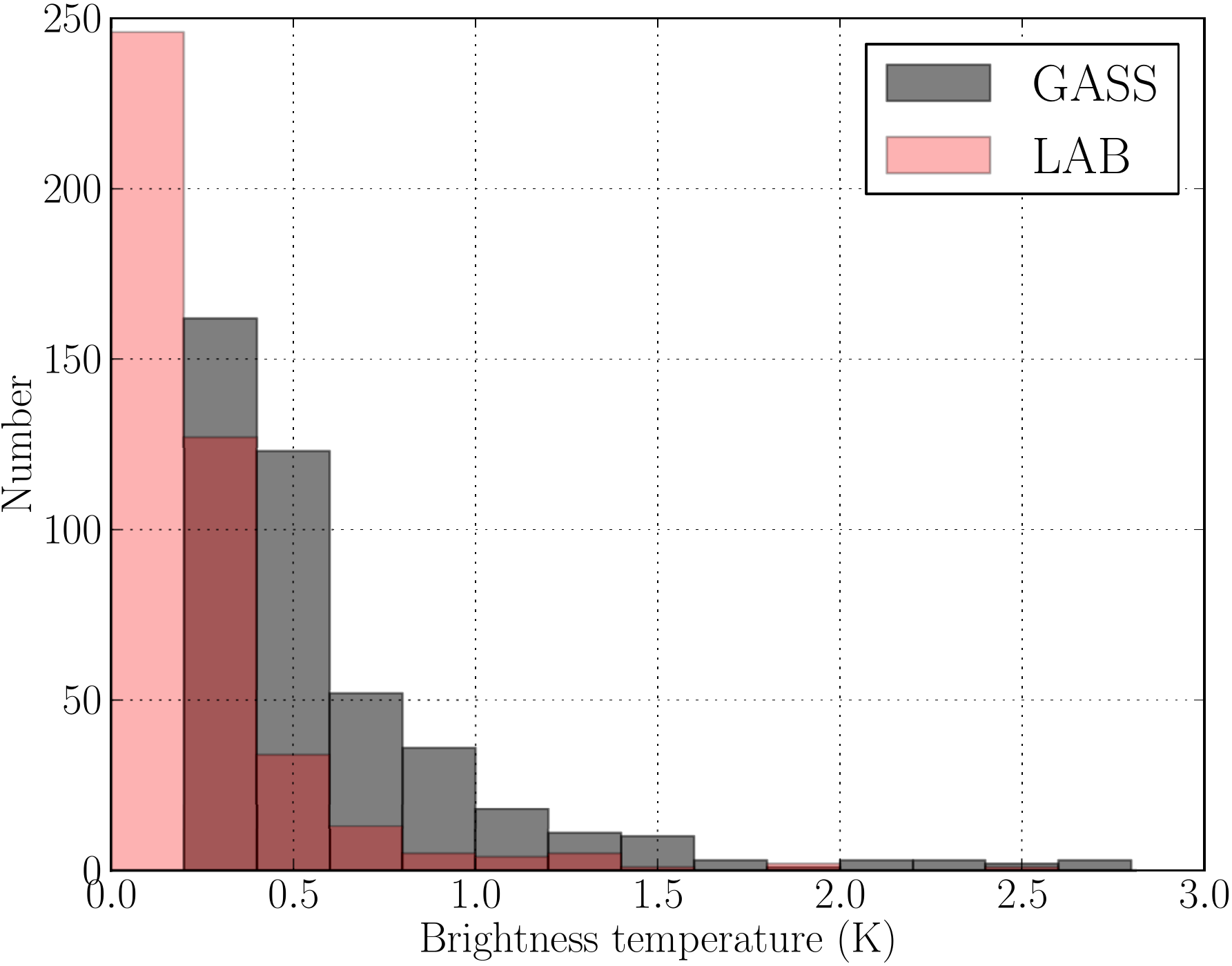}
  \caption{\small Histogram distributions of the FWHMs (left) and fitted brightness temperature peaks (right) of all 5$\sigma$ GASS clouds as measured in GASS (black) and in LAB (red). We see that the entire FWHM distribution is shifted to larger FWHMs and broadened in the case of measuring the cloud properties in LAB, while the brightness temperature distribution is shifted to lower brightness temperatures. Both of these effects are likely due to the combination of the slightly coarser spectral resolution and lower angular resolution of LAB.}
  \label{fig:labcompare2}
\end{figure*}

The distribution in FWHM of the HVC population is fairly even around the median value of 19~km~s$^{-1}$, with a slight tail towards higher FWHMs. The AVC population is clearly shifted towards lower FWHMs than the HVC population. Based on inspection of the spectra of AVCs, this shift appears to be genuine but we cannot exclude the possible effect of measuring the FWHM on partially cutoff spectra (which is the case for 213 of the 295 AVCs) leading to increased uncertainty and artificial narrowing. We find our distribution is radically different to the HIPASS HVCs,  although this is not surprising given the much coarser velocity resolution of HIPASS. In the case of HIPASS, the majority of their population was not completely resolved at the large velocity resolution of 26.4~km~s$^{-1}$. A comparison between the two catalogues is shown in Figure \ref{fig:fwhm}, with the respective velocity resolutions plotted as dashed lines. We show both the measured FWHMs of the HVC population in each survey and the FWHM distribution corrected for velocity resolution, using
\begin{equation}
\textrm{FWHM}_{observed} = \textrm{FWHM}_{actual}^2 - \Delta v^2,
\end{equation} 

which reveals good agreement between the cloud populations of GASS and HIPASS. We include in this comparison only those HIPASS clouds which are comparable with the GASS population, namely with declinations $<$~0$^\circ$ and excluding galaxies.

It is worth noting that difference between our distribution centred on a FWHM of 19~km~s$^{-1}$ is somewhat different to that found in the northern hemisphere Leiden/Dwingeloo Survey or the later Leiden/Argentina/Bonn survey data which found a distribution centred on $\sim$25-30~km~s$^{-1}$ \citep{deHeij:2002p12239,Kalberla:2006p22170}, and to the distribution found by the 140-foot telescope in Green Bank which centred on 23~km~s$^{-1}$ \citep{Cram:1976p22750}. However we note that both \citet{Cram:1976p22750} and \citet{Kalberla:2006p22170} chose to separate velocity structure into multiple spectral components, including spatial data across their sources. To better investigate our catalogue's trend towards lower FWHMs and closer match the methods of previous studies, we extracted all spectra containing at least a 5$\sigma$ signal from the spatial regions of all GASS HVCs with a total area $<$ 5~deg$^2$. This resulted in a total of 13977 spectra, of which 12180 were used as we removed any with fit LSR velocities of $<$ 90~km~s$^{-1}$. 

To determine whether a spectrum was better fit with a single component or two components, we performed fitting for both cases and chose a two-component spectrum fit if the standard deviation of the single fit residual was larger by 10\% or more than the standard deviation of the two-component fit residual. This gave a total of 13593 individual spectral components within these 12180 spectra, of which 10767 were identified as single component spectra. The results are shown in Figure \ref{fig:components}. In this case of spectral components distributed spatially across GASS HVCs, we see a median FWHM of 21~km~s$^{-1}$ ($\sigma$ = 9~km~s$^{-1}$). Compared with the result of \citet{Kalberla:2006p22170}, we find that our spectral component distribution is much narrower and concentrated on a significantly lower $\sigma$. Although there are similarities in both distributions, our narrow components, although they appear in the total distribution, are not as pronounced and we see evidence for two-component structure in the distribution of spectra identified to be single component. The incongruity is likely due both to the differing angular resolution of GASS and LAB and to the differences between their targeted population of the HVC complexes identified by \citet{Wakker:1991p22252} and our all-southern-sky population identified from GASS. 

We further investigated the lower median FWHM of the GASS HVCs by measuring their properties as found in the LAB data \citep{Kalberla:2005p22278}. We extracted the LAB spectrum closest to the best-fit position of the GASS HVCs with fitted brightness temperatures $>$ 5$\sigma$ (to aid detection) and performed the same fitting routine on each cloud. We show the results for all 5$\sigma$ and 10$\sigma$ clouds in Figure \ref{fig:labcompare}. In the case of the measured FWHMs in the LAB data, we see fairly good agreement with the GASS FWHM, although the entire distribution is positively biased. We find a median FWHM for our 5$\sigma$ clouds measured in the LAB data of $\sim$24~km~s$^{-1}$ and a mean FWHM of $\sim$28~km~s$^{-1}$. Based on these results, it appears possible that our lower median FWHM of 19~km~s$^{-1}$ compared to the previous studies of $\sim$28~km~s$^{-1}$ is a combination of the effects of slightly coarser spectral resolution (1.3~km~s$^{-1}$) and lower angular resolution (36$'$) in the LAB data rather than any significant difference from the HVC populations studied previously. We show the two distributions together in Figure \ref{fig:labcompare2}.

In fitted peak brightness temperature (Figure \ref{fig:hists2}), we see a distribution approximately following a power law, with the dashed lines on the histogram corresponding to 3$\sigma$ and 5$\sigma$. Many sources have fitted peak brightness temperatures below this cutoff because their actual peak (which was targeted during the source finding procedure) is affected by positive noise bias. This effectively increases the sensitivity of the catalogue below our target of 4$\sigma$, though the completeness is likely to be limited at lower brightness temperatures. A linear fit for both HVC and AVC distributions gives slopes of -1.4 and -0.9 respectively, with good approximation by a power law.

The column density distributions of HVCs and AVCs show an approximate power law distribution that turns over between 10$^{18}$ and 10$^{19}$~cm$^{-2}$. This suggests that we are particularly incomplete below densities of $\sim$5$\times$10$^{18}$~cm$^{-2}$. If we adopt the median FWHM of 19~km~s$^{-1}$, we see that this corresponds to an average brightness temperature of $\sim$0.15~K, which is just below the 3$\sigma$ value of 0.165~K. Based on this, we expect our catalogue to be complete to 4$\sigma$ and to closely approximate the true distribution of sources down to the 3$\sigma$ level as well.  A linear fit for both HVC and AVC distributions gives slopes of -1.3 and -1.1 respectively, with good approximation by a power law. In both cases, our fitted slopes for brightness temperature and column density are shallower than those found by \citet{Putman:2002p12244}, although closer in agreement to those of \citet{Wakker:1991p22252}. As suggested by \citet{Putman:2002p12244}, this is likely due to the fact that the GASS clouds are more similar to the complexes of clouds identified by \citet{Wakker:1991p22252}, resulting in the dominance of high brightness temperature and column density end of the distribution over the low end which becomes absorbed into the complex structure.

There is a significant spread in area amongst both AVCs and HVCs, with a few extremely large sources amongst a majority of sources below $\sim$1~deg$^2$ in size. The slight difference between the size of the clouds in the HVC population and the AVC population may indicate that the AVC population is dominated by angularly resolved sources and thus may be closer to us than those in the HVC population, although this depends considerably on the characteristic physical size of clouds and their likely distance.

\begin{figure*}
  \centering 
    \includegraphics[width=0.35\textwidth, angle=-90, trim=0 0 0 0]{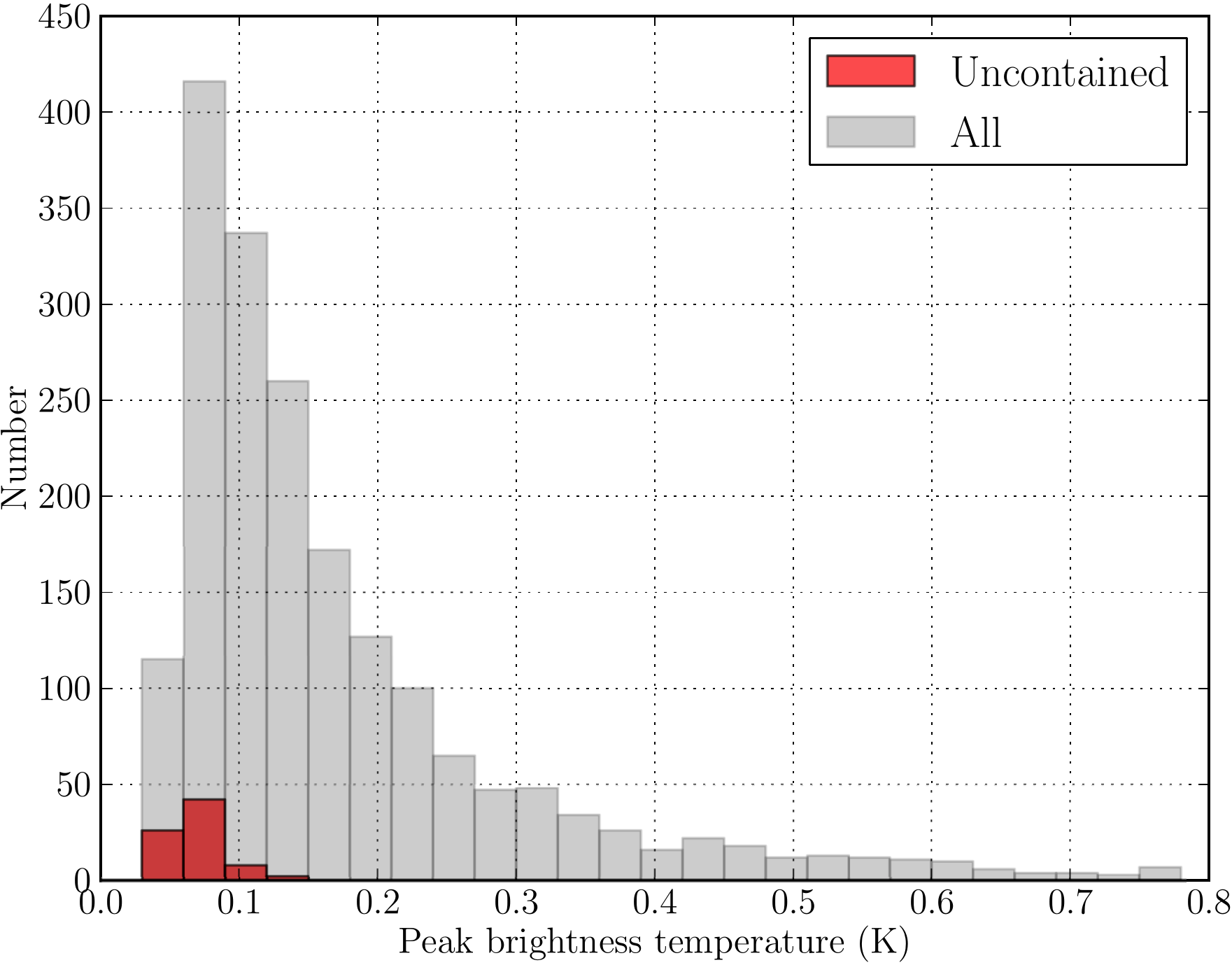}
     \includegraphics[width=0.35\textwidth, angle=-90, trim=0 0 0 0]{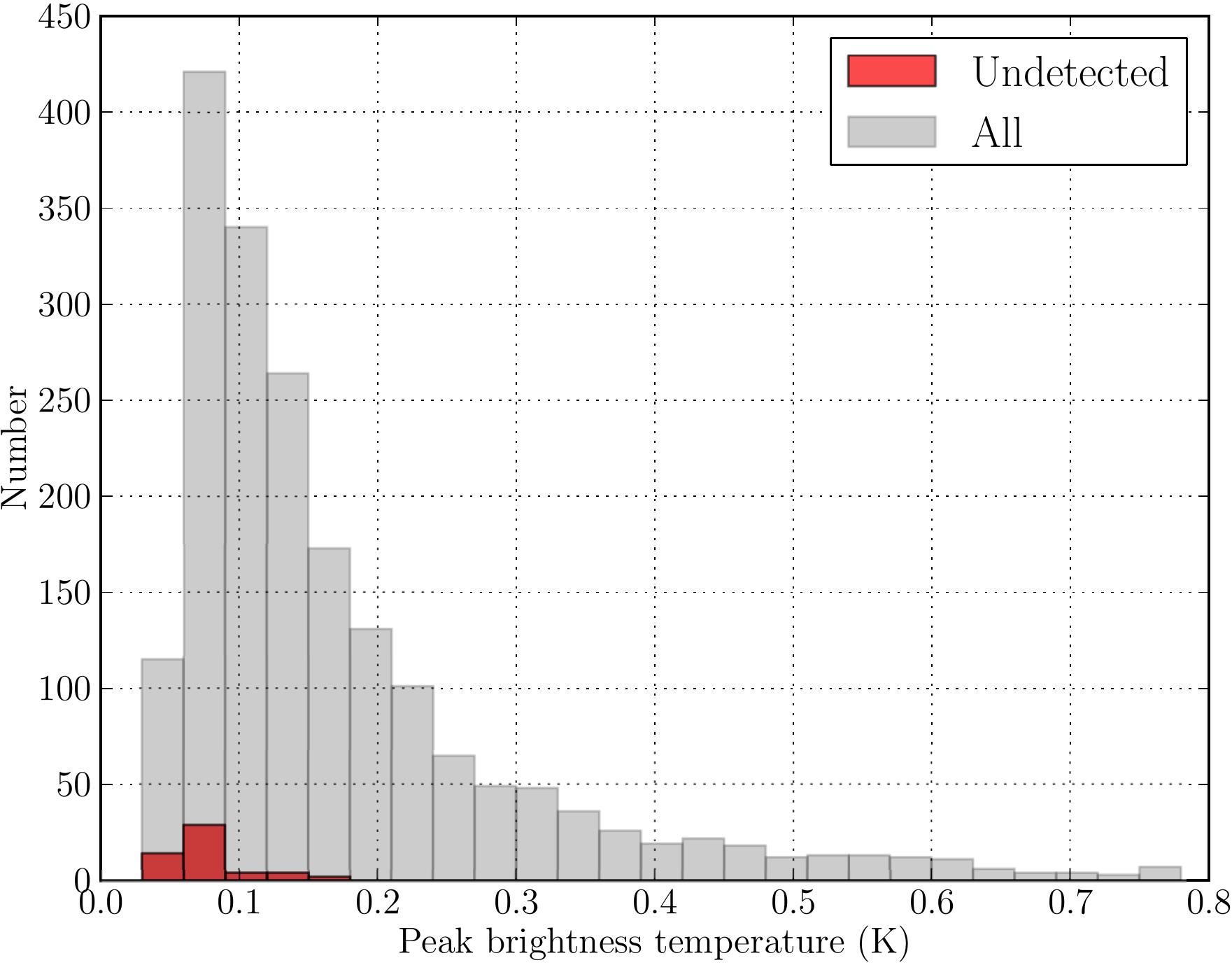}
  \caption{\small The distributions of uncontained and undetected HIPASS sources. Shown are the sources uncontained based on the intensity-weighted or peak position (left) and undetected by the KS two-sample test (right). In both cases, we find that the key limitation of the GASS HVC catalogue compared to the HIPASS catalogue is its lower sensitivity. In the case of containment, 95\% of HIPASS HVCs are contained by GASS HVCs, while in the case of detection we find that 97\% of HIPASS HVCs are detected in GASS when targeting their spectra.}
  \label{fig:undetected} 
\end{figure*}


\section{Completeness}\label{completeness}
In order to assess the completeness of our catalogue, we compare with the HIPASS HVC catalogue \citep{Putman:2002p12244}. Both catalogues were constructed using data from the Parkes 21~cm multibeam and hence share similar angular resolution (with slight differences due to the gridding process used). The key distinction between the HIPASS data and the GASS data lies in the balance between spectral resolution and sensitivity. The original spectral resolution of the HIPASS data was 18~km~s$^{-1}$ (channel resolution of 13.2~km~s$^{-1}$) with a per-channel RMS noise of $\sim$8~mK, while the GASS data has a spectral resolution of 1~km~s$^{-1}$ (channel resolution of 0.8~km~s$^{-1}$) with a per-channel RMS noise of $\sim$57~mK. Source finding for the HIPASS HVC catalogue was performed on cubes smoothed with Gaussians (31~km~s$^{-1}$, 19$^\prime$), while the subsequent measurement of clouds was performed on the Hanning-smoothed data (26.4~km~s$^{-1}$, 15.5$^\prime$). This width of $\sim$26~km~s$^{-1}$ was hence reported as the limit of resolution for the measured FWHMs of the resulting clouds. In the case of GASS, we use the limit at which we can accurately define the width of a cloud, which we assume to be three channels ($2.4$~km~s$^{-1}$). One further difference between the two data sets lies in the observing mode of HIPASS, which reduced sensitivity to large-scale structures of the order of $7^\circ$ and higher by employing the MINMED5 technique to estimate the spectral baseline within each 8 degree scan \citep{Barnes:2001p21669,Putman:2002p12244}. Conversely, GASS recovers structure of all scales by using the frequency-switching technique described in Section \ref{methodology}. As such, a direct comparison of the two catalogues is not possible.

We compare with the results of HIPASS in two ways: 1) we search for the containment of HIPASS HVCs within GASS HVCs and AVCs, and 2) we performed a targeted search of the GASS~II data for the detection of HIPASS HVCs. By containment, we mean that the HIPASS source position falls within the boundaries of a GASS cloud. By detection, we mean how many HIPASS HVCs we can detect if we look specifically in a spectrum expected to contain a signal. There are a total of 1918 HIPASS HVCs that are within the survey limits of GASS (i.e. in regions of $\alpha$, $\delta$ and $v_{LSR}$ shared by both surveys), which excludes sources with $\delta>0^\circ$ and all sources identified as galaxies. We do not exclude sources on the basis of sensitivity. Given that HIPASS scanned the sky five times versus GASS scanning the sky two times, we expect the HIPASS data to be around three times more sensitive than GASS even after the spectral resolution is accounted for. However, the exact sensitivity cutoff for GASS detection of HIPASS sources is not obvious given the differences between the data collection method, the data processing method, the pre-processing of the data prior to source finding and the actual source finding process and criteria. We thus use this analysis of completeness to provide a consideration of this sensitivity relation. The details of these comparisons are given below. 

\subsection{Containment of HIPASS HVCs by GASS HVCs and AVCs}\label{hipass_contain}
We first checked for the containment of each HIPASS source within the data cube of each GASS cloud. We used two possible positions to define the location of a HIPASS HVC: 1) the intensity-weighted position as given in the HIPASS catalogue and 2) the position of the actual peak brightness temperature in a GASS-extracted cube of the same HIPASS HVC. We included consideration of the second position because the catalogued brightness temperature and FWHM of each HIPASS HVC are measured through the peak of each source (not the intensity-weighted position). We then crossmatched both the catalogued intensity-weighted position and the position of the maximum pixel against the GASS clouds to check containment. The difference between the containment results for the two positions was two positive velocity sources and three negative velocity sources, so they were in agreement for the majority of sources.

Based on containment of the intensity-weighted and peak positions of the HIPASS sources, our catalogued clouds contain 1193 positive velocity HIPASS HVCs and 646 negative velocity HIPASS HVCs, which leaves 79 sources uncontained within the region of a GASS HVC or AVC. The distribution of uncontained sources is shown on the left (in red) in Figure \ref{fig:undetected}, with a clear limitation due to sensitivity. This method results in a containment level of $\sim$95\%, which is likely to be skewed positively due to the possibility of chance containment of sources that are not necessarily detectable. This is discussed when we crossmatch the uncontained sources against those sources undetected in Section \ref{hipass_detect}. Because the HIPASS HVC catalogue targeted compact, seemingly isolated clouds (due to the reduced sensitivity to large-scale structure), we might expect that multiple HIPASS clouds may be contained within a single GASS source rather than a one-to-one correlation. We checked this by calculating how many GASS clouds contain a HIPASS HVC, which resulted in 1230 GASS clouds containing a total of 1839 HIPASS HVCs. 

\subsection{Detection of HIPASS HVCs in GASS}\label{hipass_detect}
Another way of approaching the question of the completeness of GASS is to perform a targeted search for all HIPASS HVCs. To do this, we used the same regional definition of a HIPASS HVC as outlined in Section \ref{hipass_contain}. From each cube, we extracted two spectra. The first spectrum is taken at the reported HIPASS position, which is the intensity-weighted position of the cloud. However, because the catalogued brightness temperature and FWHM of each HIPASS HVC are measured through the peak of each source, we also extracted a spectrum at the position of the maximum pixel within the cloud.

We used the Kolgomorov-Smirnov (KS) two-sample test \citep{Smirnov:1944p23306,BolShev:1963p23312} to determine the detectability of a HIPASS HVC, with incorporated matched filtering as outlined in \citet{Jurek:2012p22797}. The KS two-sample test allows comparison of representative samples from two distributions and tests the null hypothesis that the samples were drawn from the same distribution. It returns a p-value which gives the probability of the null hypothesis being true, and as such the lower the p-value, the more different the two samples may be.

We first extracted a section of each of the two spectra of a total width that is 6 times the FWHM of the source (to ensure that we contained the source as well as local noise on either side), and then run a blind matched filtering KS test on this spectrum. We started with a window of 5 channels wide and scanned across the extracted spectrum, comparing the window with the rest of the spectrum via the KS two-sample test. We stored the position of the centre of the window as well as the window width only if the returned p-value was lower than any previous scan through the spectrum at a different window width. This then guarantees the lowest p-value for each position, with corresponding window width. The window width was widened until a maximum width of half the extracted spectrum. We selected the position with the lowest p-value (hence the greatest difference to the rest of the spectrum) as representative of the most probable location and width of any source within that spectrum.  Due to the presence of other emission in the same spectrum, we adjusted the extracted spectrum width from the default 6~$\times$~FWHM for 87 sources. 

A source was considered to be detected if the position in either spectrum of the lowest p-value (as found by the matched filtering process) fell within 1.5 times the HIPASS spectral FWHM, or within the inner 75\% of the HVC in the spectral dimension. For the 1918 sources searched we detect a total of 1864 clouds, which corresponds to a detection level of just over 97\%. The undetected clouds are shown on the right (in red) in Figure \ref{fig:undetected}, with again the key limitation being sensitivity. When comparing these results to the results obtained in Section \ref{hipass_contain}, we find that we have slightly higher completeness using this method, which makes sense in the context of being able to detect a source given a known position versus ability to source-find blindly on data.

When we crossmatch the undetected clouds against the uncontained clouds, we find only 6 clouds overlap in both samples. This result suggests that the HIPASS sources that are uncontained by GASS HVCs are not uncontained because of undetectability. The fact that they are detected by the KS test and yet do not have their intensity-weighted or peak positions contained by any GASS HVCs indicates that their peaks are still too low compared with our source finding criteria, but their signatures are still present in the data however faintly. This is not entirely surprising given that HIPASS is more sensitive than GASS. Conversely, some undetected HIPASS clouds are likely to be contained by chance coincidence within GASS HVCs which is why we see some undetected clouds that have been classified as contained. 

In the context of the distinction between which HIPASS sources are successfully `found' within the GASS HVCs versus those which are detectable when we specifically target their region, we note that the KS test method of searching is not an unbiased, blind search for sources. For this case, we targeted known sources with known positions, and as such are source finding with prior knowledge that there should be a signal present. This prior knowledge is key to why we detect more sources that we contain, because it gives us the ability to differentiate between regions of increased noise, artifacts, interference and actual source signal. Because of this we do not expect to have actually found all of the HIPASS sources (even those that are detectable) with our blind source finder. It is unlikely that a different source finder would do significantly better because source finding is limited by differentiation between other aspects of the data that look like source signal but are in fact spurious. 

\subsubsection{Summary of completeness and reliability}
We have used two methods to assess the completeness of the GASS HVC catalogue: 1) the containment of HIPASS HVCs and 2) the detection of HIPASS HVCs. We find that the GASS HVCs and AVCs contain 95\% of the HIPASS HVCs despite the slightly better sensitivity of the HIPASS data, although this number likely contains a small number of coincidently contained sources beyond our sensitivity limit. Using a blind matched filtering script at the spatial position of each HIPASS HVC, we are able to detect 97\% of the HIPASS population. Based on these results, we conclude that the completeness of the GASS HVC catalogue is at least 95\% compared with the HIPASS HVCs (which were found with more sensitive data), and at our 4$\sigma$ cutoff of 0.22~K it should be close to 100\% complete. Due to the manual inspection of candidate sources described in Section \ref{dataproc}, the catalogue will be almost 100\% reliable.


\section{Summary and future work}\label{conclusion}

We have presented a catalogue of 1693 high velocity clouds (HVCs) as found in the Galactic All Sky Survey data, which has a sensitivity of $\sim$57~mK, an angular resolution of $\sim$16$'$ and a spectral resolution of $\sim$1~km~s$^{-1}$. These clouds have been identified using the GASS~II data, which includes stray-radiation correction. Our catalogue contains clouds down to peak brightness temperatures of 4$\sigma$ ($\sim$230~mK) at all declinations $\delta$ $< 0^\circ$. There are a total of 1111 positive velocity clouds and 582 negative velocity clouds, distributed inhomogeneously across the southern sky with concentrations correlated with the Magellanic system and with the Galactic disk.  Of our sample of HVCs (excluding AVCs), we find that 567 clouds were previously uncatalogued by the HIPASS survey and as such are new in the context of southern sky surveys for HVCs. A total of 758 clouds in our catalogue have no HIPASS counterpart and do not contain a HIPASS cloud in their region. Our final catalogue of GASS HVCs has a reliability close to 100\% given that each source has been manually inspected, as well as being complete to the 4$\sigma$ level within the constraints of our search parameters and masking procedure. In order to find any narrow clouds potentially missed by searching on binned data, we performed a targeted narrow cloud search on the unbinned data. This resulted in adding only 14 narrow clouds to the catalogue, which may suggest that isolated narrow-line clouds (without a co-located broad component) are intrinsically rare or possibly that they are rare at the $\sim$16$'$ resolution of GASS. 

Our catalogue also includes  295 anomalous velocity clouds (AVCs) which do not meet the velocity criteria associated with the definition of an HVC although they do deviate from Galactic rotation. This sample of AVCs includes 197 positive velocity clouds and 98 negative velocity clouds. As many of these AVCs skirt the tenuous boundary between IVC and HVC, we designate them as AVC pending further investigation. Eventually separating different populations of clouds on the basis of their physical properties, which should correlate with their dynamical properties, may help constrain the current definition of HVCs and help separate them from gas that is more clearly associated with the rotating disk as well as mitigate the effects of line-of-sight projection.

We have investigated the overall properties of the GASS HVC and AVC population and compared these with the results of previous surveys. There is an apparent preference towards negative velocities for our clouds in the Galactic Standard of Rest frame, which is in agreement with previous findings on the high-velocity sky \citep{Stark:1992p23268,Haffner:2003p20906}. We find that our median FWHM of 19~km~s$^{-1}$ is lower than that of other studies, and a two-component spectral analysis similar to that of \citet{Kalberla:2006p22170} reveals that our distribution of spectral components is similarly centred on a lower FWHM than previous studies. This appears to be predominantly a function of the differing angular and spectral resolution between surveys, but may also be suggestive about the properties of the southern sky distribution of HVCs. Our FWHM distribution compared to that of HIPASS is significantly shifted to lower velocities, indicating that we are probing the true physical distribution of HVC FWHMs without spectral resolution limitations. 

We have compared our catalogue with the previous deep HIPASS data in order to judge the completeness of the GASS HVC population. We find that our HVCs contain 95\% of the peaks of all HIPASS HVCs. When running a blind matched filtering window on each HIPASS cloud, we find that we are able to detect over 97\% of these HVCs within the GASS~II data. However we find little overlap in the uncontained versus undetected samples, which is likely due to a combination of chance containment of sources within large HVC complexes and detection limits existing below source finding limits. 

The combination of high angular resolution and high sensitivity with enhanced image fidelity (via both frequency-switching and stray radiation correction) in an all-sky data set make GASS an excellent means of seeing the large-scale structure of the gaseous halo as well as the composition at smaller scales. The spectral resolution of the GASS data gives us an unprecedented look into the complicated velocity structure of HVCs; we see evidence for multiple spectral components, asymmetries and complex spectral structure. These HVCs are not isolated objects, as they are in the process of interacting with both the gravitational field of the Milky Way and with the high temperature environment of the halo medium. Our upcoming work on the velocity structure of GASS HVCs and AVCs (Moss et al, in preparation) will target clouds which demonstrate head-tail structure, complex spectral structure or extremely narrow line widths, in order to probe the conditions of clouds which are interacting directly with their surrounding medium and how this relates to disk-halo interaction in the Milky Way. The GASS HVC catalogue combined with the GASS data from which it was derived will be useful to any work which seeks to investigate how gas is infalling into and outflowing out of the Milky Way both morphologically and dynamically, on the largest scales as well as in the fine detail revealed by the high angular and spectral resolution of GASS. 


\acknowledgments
We would like to acknowledge our anonymous referee for their constructive suggestions, A.J. Green and A.J. Walsh for useful discussions during the development of this paper, R. Jurek for assistance with the application of his blind matched filtering detection algorithm and C.R. Purcell for help with the manuscript format. This research made use of APLpy, an open-source plotting package for Python hosted at \underline{\url{http://aplpy.github.com}}.




\bibliographystyle{apj}
\bibliography{all4}

\begin{thebibliography}{43}
\expandafter\ifx\csname natexlab\endcsname\relax\def\natexlab#1{#1}\fi

\bibitem[{Barnes {et~al.}(2001)Barnes, Staveley-Smith, de~Blok, Oosterloo,
  Stewart, Wright, Banks, Bhathal, Boyce, Calabretta, Disney, Drinkwater,
  Ekers, Freeman, Gibson, Green, Haynes, Hekkert, Henning, Jerjen, Juraszek,
  Kesteven, Kilborn, Knezek, Koribalski, Kraan-Korteweg, Malin, Marquarding,
  Minchin, Mould, Price, Putman, Ryder, Sadler, Schr{\"o}der, Stootman,
  Webster, Wilson, \& Ye}]{Barnes:2001p21669}
Barnes, D.~G., Staveley-Smith, L., de~Blok, W. J.~G., {et~al.} 2001, MNRAS,
  322, 486

\bibitem[{Bekhti {et~al.}(2006)Bekhti, Br{\"u}ns, Kerp, \&
  Westmeier}]{BenBekhti:2006p22800}
Bekhti, N.~B., Br{\"u}ns, C., Kerp, J., \& Westmeier, T. 2006, A{\&}A, 457, 917

\bibitem[{Bol'Shev(1963)}]{BolShev:1963p23312}
Bol'Shev, L. 1963, Theory of Probability {\&} Its Applications, 8, 121

\bibitem[{Boomsma {et~al.}(2008)Boomsma, Oosterloo, Fraternali, van~der Hulst,
  \& Sancisi}]{Boomsma:2008p23611}
Boomsma, R., Oosterloo, T.~A., Fraternali, F., van~der Hulst, J.~M., \&
  Sancisi, R. 2008, A{\&}A, 490, 555

\bibitem[{Braun \& Burton(1999)}]{Braun:1999p22713}
Braun, R., \& Burton, W.~B. 1999, A{\&}A, 341, 437

\bibitem[{Br{\"u}ns \& Westmeier(2004)}]{Bruns:2004p20402}
Br{\"u}ns, C., \& Westmeier, T. 2004, A{\&}A, 426, L9

\bibitem[{Cram \& Giovanelli(1976)}]{Cram:1976p22750}
Cram, T.~R., \& Giovanelli, R. 1976, A{\&}A, 48, 39

\bibitem[{de~Heij {et~al.}(2002)de~Heij, Braun, \& Burton}]{deHeij:2002p12239}
de~Heij, V., Braun, R., \& Burton, W.~B. 2002, A{\&}A, 391, 159

\bibitem[{Giovanelli {et~al.}(2010)Giovanelli, Haynes, Kent, \&
  Adams}]{Giovanelli:2010p23648}
Giovanelli, R., Haynes, M.~P., Kent, B.~R., \& Adams, E. A.~K. 2010, ApJL, 708,
  L22

\bibitem[{Giovanelli {et~al.}(1973)Giovanelli, Verschuur, \&
  Cram}]{Giovanelli:1973p23469}
Giovanelli, R., Verschuur, G.~L., \& Cram, T.~R. 1973, A{\&}AS, 12, 209

\bibitem[{Haffner {et~al.}(2003)Haffner, Reynolds, Tufte, Madsen, Jaehnig, \&
  Percival}]{Haffner:2003p20906}
Haffner, L.~M., Reynolds, R.~J., Tufte, S.~L., {et~al.} 2003, ApJS, 149, 405

\bibitem[{Hartmann \& Burton(1997)}]{Hartmann:1997p10021}
Hartmann, D., \& Burton, W.~B. 1997, Atlas of Galactic Neutral Hydrogen, iSBN:
  0521471117

\bibitem[{Heald {et~al.}(2011)Heald, Allan, Zschaechner, Kamphuis, Rand,
  J{\'o}zsa, \& Gentile}]{Heald:2011p23622}
Heald, G., Allan, J., Zschaechner, L., {et~al.} 2011, Tracing the Ancestry of
  Galaxies (on the land of our ancestors), 277, 59

\bibitem[{Hulsbosch(1978)}]{Hulsbosch:1978p23464}
Hulsbosch, A. N.~M. 1978, A{\&}A, 66, L5

\bibitem[{Jurek(2012)}]{Jurek:2012p22797}
Jurek, R. 2012, PASA, 29, 251

\bibitem[{Kalberla {et~al.}(2005)Kalberla, Burton, Hartmann, Arnal, Bajaja,
  Morras, \& P{\"o}ppel}]{Kalberla:2005p22278}
Kalberla, P. M.~W., Burton, W.~B., Hartmann, D., {et~al.} 2005, A{\&}A, 440,
  775

\bibitem[{Kalberla \& Haud(2006)}]{Kalberla:2006p22170}
Kalberla, P. M.~W., \& Haud, U. 2006, A{\&}A, 455, 481

\bibitem[{Kalberla {et~al.}(2010)Kalberla, McClure-Griffiths, Pisano,
  Calabretta, Ford, Lockman, Staveley-Smith, Kerp, Winkel, Murphy, \&
  Newton-McGee}]{Kalberla:2010p19880}
Kalberla, P. M.~W., McClure-Griffiths, N.~M., Pisano, D.~J., {et~al.} 2010,
  A{\&}A, 521, 17

\bibitem[{Kamphuis {et~al.}(2011)Kamphuis, Peletier, van~der Kruit, \&
  Heald}]{Kamphuis:2011p23605}
Kamphuis, P., Peletier, R.~F., van~der Kruit, P.~C., \& Heald, G.~H. 2011,
  MNRAS, 414, 3444

\bibitem[{Lehner \& Howk(2010)}]{Lehner:2010p22602}
Lehner, N., \& Howk, J.~C. 2010, ApJL, 709, L138

\bibitem[{Lehner {et~al.}(2009)Lehner, Staveley-Smith, \&
  Howk}]{Lehner:2009p22600}
Lehner, N., Staveley-Smith, L., \& Howk, J.~C. 2009, ApJ, 702, 940

\bibitem[{McClure-Griffiths {et~al.}(2009)McClure-Griffiths, Pisano,
  Calabretta, Ford, Lockman, Staveley-Smith, Kalberla, Bailin, Dedes,
  Janowiecki, Gibson, Murphy, Nakanishi, \&
  Newton-McGee}]{McClureGriffiths:2009p3462}
McClure-Griffiths, N.~M., Pisano, D.~J., Calabretta, M.~R., {et~al.} 2009,
  ApJS, 181, 398

\bibitem[{Muller {et~al.}(1963)Muller, Oort, \& Raimond}]{Muller:1963p22387}
Muller, C.~A., Oort, J.~H., \& Raimond, E. 1963, Comptes Rendus l'Academie des
  Sciences, 257, 1661

\bibitem[{Murphy {et~al.}(2007)Murphy, Mauch, Green, Hunstead, Piestrzynska,
  Kels, \& Sztajer}]{Murphy2007a}
Murphy, T., Mauch, T., Green, A., {et~al.} 2007, MNRAS, 382, 382

\bibitem[{Peek {et~al.}(2009)Peek, Heiles, Putman, \&
  Douglas}]{Peek:2009p16944}
Peek, J., Heiles, C., Putman, M., \& Douglas, K. 2009, ApJ, 692, 827

\bibitem[{Peek {et~al.}(2012)Peek, Heiles, Douglas, Lee, Grcevich,
  Stanimirovi{\'c}, Putman, Korpela, Gibson, Begum, Saul, Robishaw, \& Kr{\v
  c}o}]{Peek:2012p22763}
Peek, J. E.~G., Heiles, C., Douglas, K.~A., {et~al.} 2012, ApJS, 194, 20

\bibitem[{Putman {et~al.}(2002)Putman, de~Heij, Staveley-Smith, Braun, Freeman,
  Gibson, Burton, Barnes, Banks, Bhathal, de~Blok, Boyce, Disney, Drinkwater,
  Ekers, Henning, Jerjen, Kilborn, Knezek, Koribalski, Malin, Marquarding,
  Minchin, Mould, Oosterloo, Price, Ryder, Sadler, Stewart, Stootman, Webster,
  \& Wright}]{Putman:2002p12244}
Putman, M.~E., de~Heij, V., Staveley-Smith, L., {et~al.} 2002, AJ, 123, 873

\bibitem[{Sancisi {et~al.}(2008)Sancisi, Fraternali, Oosterloo, \& van~der
  Hulst}]{Sancisi:2008p23612}
Sancisi, R., Fraternali, F., Oosterloo, T., \& van~der Hulst, T. 2008,
  A{\&}ARv, 15, 189

\bibitem[{Saul {et~al.}(2012)Saul, Peek, Grcevich, Putman, Douglas, Korpela,
  Stanimirovi{\'c}, Heiles, Gibson, Lee, Begum, Brown, Burkhart, Hamden,
  Pingel, \& Tonnesen}]{Saul:2012p22666}
Saul, D.~R., Peek, J. E.~G., Grcevich, J., {et~al.} 2012, ApJ, 758, 44

\bibitem[{Sembach {et~al.}(2002)Sembach, Gibson, Fenner, \&
  Putman}]{Sembach:2002p22598}
Sembach, K.~R., Gibson, B.~K., Fenner, Y., \& Putman, M.~E. 2002, ApJ, 572, 178

\bibitem[{Smirnov(1944)}]{Smirnov:1944p23306}
Smirnov, N. 1944, Uspekhi Matematicheskikh Nauk, 10, 179

\bibitem[{Smoker {et~al.}(2011)Smoker, Fox, \& Keenan}]{Smoker:2011p22447}
Smoker, J.~V., Fox, A.~J., \& Keenan, F.~P. 2011, MNRAS, 415, 1105

\bibitem[{Stark {et~al.}(1992)Stark, Gammie, Wilson, Bally, Linke, Heiles, \&
  Hurwitz}]{Stark:1992p23268}
Stark, A.~A., Gammie, C.~F., Wilson, R.~W., {et~al.} 1992, ApJS, 79, 77

\bibitem[{Taylor {et~al.}(1999)Taylor, Carilli, \& Perley}]{Taylor:1999p23578}
Taylor, G.~B., Carilli, C.~L., \& Perley, R.~A. 1999, Synthesis Imaging in
  Radio Astronomy II, 180

\bibitem[{Tenorio-Tagle \& Bodenheimer(1988)}]{TenorioTagle:1988p13608}
Tenorio-Tagle, G., \& Bodenheimer, P. 1988, ARA{\&}A, 26, 145

\bibitem[{Tripp \& Song(2012)}]{Tripp:2012p22503}
Tripp, T.~M., \& Song, L. 2012, ApJ, 746, 173

\bibitem[{Wakker(1991)}]{Wakker:1991p22250}
Wakker, B.~P. 1991, A{\&}A, 250, 499

\bibitem[{Wakker {et~al.}(2002)Wakker, Oosterloo, \&
  Putman}]{Wakker:2002p22578}
Wakker, B.~P., Oosterloo, T.~A., \& Putman, M.~E. 2002, AJ, 123, 1953

\bibitem[{Wakker \& van Woerden(1991)}]{Wakker:1991p22252}
Wakker, B.~P., \& van Woerden, H. 1991, A{\&}A, 250, 509

\bibitem[{Wakker \& van Woerden(1997)}]{Wakker:1997p9859}
---. 1997, ARA{\&}A, 35, 217

\bibitem[{Wannier {et~al.}(1972)Wannier, Wrixon, \&
  Wilson}]{Wannier:1972p23447}
Wannier, P., Wrixon, G.~T., \& Wilson, R.~W. 1972, A{\&}A, 18, 224

\bibitem[{Westmeier {et~al.}(2007)Westmeier, Braun, Br{\"u}ns, Kerp, \&
  Thilker}]{Westmeier:2007p23592}
Westmeier, T., Braun, R., Br{\"u}ns, C., Kerp, J., \& Thilker, D.~A. 2007, New
  Astronomy Reviews, 51, 108

\bibitem[{Westmeier {et~al.}(2005)Westmeier, Braun, \&
  Thilker}]{Westmeier:2005p23593}
Westmeier, T., Braun, R., \& Thilker, D. 2005, A{\&}A, 436, 101

\end{thebibliography}

\end{document}